\documentclass[10pt,final,journal]{IEEEtran}

\usepackage{graphicx}
\graphicspath{{Figures/}}
\usepackage{float}
\usepackage{array}
\usepackage{mathtools}
\usepackage{amsmath} 
\usepackage{amssymb}
\usepackage{amsxtra}
\usepackage{amstext}
\usepackage{dsfont}
\usepackage{bbm}
\usepackage{latexsym}
\usepackage{epsfig}
\usepackage{epstopdf}
\usepackage{bbm}
\usepackage{enumitem}
\usepackage{multirow}
\usepackage{diagbox}
\usepackage{slashbox}
\usepackage{algorithm}
\usepackage{algpseudocode}
\usepackage{longtable}

\makeatletter
\def\BState{\State\hskip-\ALG@thistlm}
\makeatother

\usepackage{caption}
\usepackage{cite}
\usepackage[T1]{fontenc}
\usepackage{setspace}
\usepackage{lscape}
\usepackage[cmintegrals]{newtxmath}
\usepackage{url}
\usepackage{tabularx,colortbl}
\usepackage[table]{xcolor}
\usepackage{units}
\usepackage{cases}
\usepackage{hyperref}
\usepackage[ subrefformat=parens,labelformat=parens, caption=false, font=footnotesize]{subfig}

\usepackage{color}

\newcommand{\nth}[1]{{#1}{\text{th}}}
\newcommand{\abs}[1]{\left|{#1}\right|}
\newcommand{\norm}[1]{\left\|{#1}\right\|}
\newcommand{\invb}[1]{\mathbf{#1}^\mathrm{-1}}

\newcommand{\tranb}[1]{\mathbf{#1}^\mathrm{T}}
\newcommand{\mbf}[1]{\mathbf{#1}}
\newcommand{\Hr}{\mathcal{H}}
\newcommand{\Tr}{\mathrm{T}}

\newcommand{\atantwo}{\mathrm{arctan2}}

\newcommand{\clu}{\mathrm{clu}}
\newcommand{\ray}{\mathrm{ray}}
\newcommand{\LoS}{\mathrm{LoS}}
\newcommand{\NLoS}{\mathrm{NLoS}}

\newcommand{\GHz}{\mathrm{GHz}}

\newcommand{\svv}{\mathrm{sv}}
\newcommand{\spp}{\mathrm{spp}}

\newcommand{\coh}{\mathrm{coh}}
\newcommand{\rms}{\mathrm{rms}}
\newcommand{\sym}{\mathrm{sym}}
\newcommand{\maxx}{\mathrm{dmax}}
\newcommand{\sig}{\mathrm{sig}}
\newcommand{\RF}{\mathrm{RF}}
\newcommand{\BBand}{\mathrm{BB}}

\newcommand{\cent}{\mathrm{cen}}
\newcommand{\GMM}{\mathrm{GMM}}
\newcommand{\subb}{\mathrm{sub}}
\newcommand{\htwoo}{\mathrm{H2O}}

\newcommand{\apx}{\mathrm{aprx}}
\newcommand{\Jakes}{\mathrm{Jakes}}
\newcommand{\Flat}{\mathrm{Flat}}

\long\def\comment#1{}

\newfont{\bbb}{msbm10 scaled 700}

\newfont{\bb}{msbm10 scaled 1100}
\newcommand{\CC}{\mbox{\bb C}}
\newcommand{\PP}{\mbox{\bb P}}
\newcommand{\RR}{\mbox{\bb R}}

\newcommand{\EE}{\mbox{\bb E}}

\makeatletter
\def\ps@IEEEtitlepagestyle{%
  \def\@oddfoot{\mycopyrightnotice}%
  \def\@oddhead{\hbox{}\@IEEEheaderstyle\leftmark\hfil\thepage}\relax
  \def\@evenhead{\@IEEEheaderstyle\thepage\hfil\leftmark\hbox{}}\relax
  \def\@evenfoot{}%
}

\def\mycopyrightnotice{%
  \begin{minipage}{\textwidth}
  \centering \scriptsize
  Copyright~\copyright~2021 IEEE. Personal use of this material is permitted. Permission from IEEE must be obtained for all other uses, in any current or future media, including\\reprinting/republishing this material for advertising or promotional purposes, creating new collective works, for resale or redistribution to servers or lists, or reuse of any copyrighted component of this work in other works by sending a request to pubs-permissions@ieee.org.
  \end{minipage}
}
\makeatother

\begin{document}

\title{TeraMIMO: A Channel Simulator for Wideband Ultra-Massive MIMO Terahertz Communications}

\author{Simon~Tarboush,
        Hadi~Sarieddeen,~\IEEEmembership{Member,~IEEE,}
        Hui~Chen,~\IEEEmembership{Member,~IEEE,}
        Mohamed~Habib~Loukil,
        Hakim~Jemaa,
        Mohamed-Slim~Alouini,~\IEEEmembership{Fellow,~IEEE,}
        and Tareq~Y.~Al-Naffouri,~\IEEEmembership{Senior Member,~IEEE}
\thanks{This work was supported by the KAUST Office of Sponsored Research. The MATLAB package and user interface of TeraMIMO are available on GitHub: \url{https://github.com/hasarieddeen/TeraMIMO}. S. Tarboush is from Damascus, Syria (e-mail: simon.w.tarboush@gmail.com). The rest of the authors are with the Department of Computer, Electrical and Mathematical Sciences and Engineering (CEMSE), King Abdullah University of Science and Technology (KAUST), Thuwal, Makkah Province, Kingdom of Saudi Arabia, 23955-6900 (e-mail: hadi.sarieddeen@kaust.edu.sa; hui.chen@kaust.edu.sa; mohamedhabib.loukil@kaust.edu.sa; hakim.jemaa@kaust.edu.sa; slim.alouini@kaust.edu.sa; tareq.alnaffouri@kaust.edu.sa).}
}

\maketitle
\IEEEpubidadjcol
\begin{abstract}

Following recent advances in terahertz (THz) technology, there is a consensus on the crucial role of THz communications in the next generation of wireless systems.
Aiming at catalyzing THz communications research, we propose TeraMIMO, an accurate stochastic MATLAB simulator of statistical THz channels.
We simulate ultra-massive multiple-input multiple-output antenna configurations as critical infrastructure enablers that overcome the limitation in THz communications distances.
We consider both line-of-sight and multipath components and propose frequency- and delay-domain implementations for single- and multi-carrier paradigms in both time-invariant and time-variant scenarios.
We implement exhaustive molecular absorption computations based on radiative transfer theory alongside alternative sub-THz approximations. We further model THz-specific constraints, including wideband beam split effects, spherical wave propagation, and misalignment fading.
We verify TeraMIMO by analogy with measurement-based channel models in the literature and ergodic capacity analysis.
We introduce a graphical user interface and a guide for using TeraMIMO in THz channel generation and analyses.
\end{abstract}

\maketitle
\section{Introduction}
\label{sec:intro}

\IEEEPARstart{T}{he} increased interest in the unutilized part of the radio-frequency (RF) spectrum over carrier frequencies from 0.1 terahertz (THz) to $\unit[10]{THz}$ \cite{kurner2014towards,Akyildiz6882305} follows recent electronic, photonic, and plasmonic advancements that have enabled efficient THz signal generation, radiation, and modulation \cite{sengupta2018terahertz,6708549Jornet}. The THz band promises Terabit-per-second (Tbps) data rates and massive secure connectivity in terrestrial, aerial, and satellite networks, with novel sensing, imaging, and localization capabilities that could shape the sixth-generation (6G) of wireless communication systems and beyond \cite{zhang20196g,sarieddeen2019generation,rajatheva2020scoring,Rappaport8732419,de2021convergent,you2021towards}.

In addition to the existing challenges on the device level, several challenges on the infrastructure and algorithmic levels have to be addressed to realize THz communications. For instance, THz signals are subject to high propagation losses that limit the communication distances. Towards extending the THz communication range, ultra-massive multiple-input multiple-output (UM-MIMO) antenna arrays \cite{Sarieddeen2019SM,akyildiz2016realizing,Han2018UMMIMO} and intelligent reflecting surfaces (IRSs) \cite{ma2020intelligent,faisal2019ultra} are two crucial infrastructure enablers. Furthermore, THz-specific signal processing \cite{sarieddeen2021overview} and networking \cite{Ghafoor9170557} algorithms are vital for mitigating the quasi-optical THz propagation constraints. For instance, although molecular absorption results in spectrum shrinking, distance-adaptive resource allocation \cite{Han2016M-W-D-A,Han2016_DABARA} can maintain efficient spectrum usage.
Optimized THz single-carrier (SC) and multi-carrier (MC) modulations can also enable joint communication and sensing applications.
However, accurate knowledge of the THz channel is a prerequisite for efficient THz-specific signal processing~\cite{sarieddeen2021overview,Weithoffer8109974}.

Our knowledge on THz channel models and characteristics is still developing at this early stage of THz research~\cite{wang20206g}, where ray-tracing (RT)-based and measurement campaigns continue to report new findings for both sub-THz and THz bands.
RT techniques result in accurate channel models but require exact information about the propagation environment and material properties.
Several RT-based THz channel models are reported in the literature. A unified multi-ray model (experimentally validated over $\unit[0.06-1]{THz}$) is presented in \cite{Han2015MR}, accounting for the line-of-sight (LoS), reflected, diffracted, and scattered paths. A deterministic indoor channel model is developed in \cite{Moldovan2014}, based on both Kirchhoff scattering theory and RT simulations, in the frequency range $\unit[0.1-1]{THz}$. Also, three-dimensional (3D) end-to-end channel models are developed in \cite{han_channel_2016,Han2018UMMIMO} by incorporating Graphene-based antennas. Other recently reported indoor LoS THz channel measurements include the works in \cite{Abbasi2020} for three bands of $\unit[10]{GHz}$ bandwidth over $\unit[140-220]{GHz}$, and in \cite{Serghiou2020} for ultra-wideband channels of $\unit[250]{GHz}$ bandwidth over $\unit[500-750]{GHz}$. Furthermore, sub-THz ($\unit[142]{GHz}$) outdoor urban ($\unit[120]{m}$) channel models for both LoS and non line-of-sight (NLoS) scenarios are reported in \cite{xing2021propagation}; the same models are verified for indoor scenarios in \cite{xing2021millimeter}. RT simulations are compared with channel measurements in \cite{Priebe2010_300ghz_10bw} for a $\unit[0.3]{THz}$ ($\unit[10]{GHz}$ bandwidth) indoor scenario, whereas in \cite{Priebe2013_300ghz_50bw}, an extensive measurement campaign is conducted for $2\!\times\!2$ MIMO indoor wideband channels over $\unit[275-325]{GHz}$ ($\unit[50]{GHz}$ bandwidths). Further channel measurements are reported at $\unit[300]{GHz}$ in \cite{Priebe2011,Ostmann2012}.

On the other hand, statistical channel modeling is widely applicable, requires less geometric information, and has low computational complexity; however, it suffers from low accuracy \cite{Han_channelmodels}. Statistical parameters for short–range LoS scenarios over $\unit[0.24-0.3]{THz}$ are derived in \cite{ekti2017statistical}. A stochastic spatio-temporal $\unit[0.3]{THz}$ indoor channel model is also introduced in \cite{priebe2013stochasticchannelmodel} using a modified Saleh-Valenzuela (SV) model. Major modifications to the THz-specific SV model are then introduced in \cite{lin2015indoor}, especially in modeling the multipath (MP) parameters such as angles of departure/arrival (AoDs/AoAs) and time of arrival (ToA). Another variation of a cluster-based THz SV channel model that modifies the power delay profile distributions is proposed in \cite{Cheng_datacenter_2020} for a data center environment at $\unit[300]{GHz}$. Moreover, in \cite{zhao2019extending}, an extension to spatial and temporal characterizations in a typical indoor THz environment is presented by comparing channel measurement results at $\unit[0.35]{THz}$ and $\unit[0.65]{THz}$. Other recent statistical THz channel modeling efforts include the work in \cite{tekbiyik2021modeling}, which examines a Gamma mixture model over measurements in the $\unit[240-300]{GHz}$ range, and the work in \cite{Wu2021}, which characterizes the indoor environment in the presence of severe reflection losses, MP fading, and indoor blockage effects. Although a comparison with a multi-ray RT THz channel model is performed in the latter, the lack of measurement data still limits the validation process over the THz-bands.
A measurement-based indoor wideband $\unit[130-143]{GHz}$ channel model is investigated in \cite{Yu2020} using RT techniques to post-process measured data. Several channel characteristics are derived, such as temporal and spatial sparsity, where the number of MP components is found to be less than ten. An extension in \cite{chen2021channel} introduces a hybrid indoor THz channel model combining both RT and statistical methods. In \cite{s21062015}, measurement-based sub-THz ($\unit[107-109]{GHz}$) channel characteristics are captured for industrial environments.

Novel reliable high-speed communications such as vehicle-to-everything (V2X), vehicle-to-vehicle (V2V), ultra-high-speed smart rail mobility, and drones communications are also expected to be pivotal for 6G vertical industries. Such scenarios are challenging for 6G channel modeling because of the resultant non-stationary time-variant (TV) channels and the underlying statistical properties~\cite{jiang2020channel,wang20206g}.
Combined with large bandwidths and UM-MIMO arrays, high-speed mobility results in non-stationary characteristics in the space, time, and frequency domains \cite{paier2008non,yuan20153d}.
Many models are proposed in the literature for such channels \cite{yuan20153d,jiang20193}, typically geometry-based stochastic models (GBSMs) and their variants. A $\unit[110]{GHz}$ urban vehicle-to-infrastructure channel model is analyzed using RT techniques  in~\cite{chen2019time}; a 3D THz indoor non-stationary space-time-frequency massive MIMO channel model is proposed and verified theoretically and empirically in~\cite{wang2020novel}; THz channel models for smart rail mobility scenarios are proposed and studied via RT simulations in~\cite{guan2016millimeter,guan2021channel}; measurement campaigns for $\unit[300]{GHz}$ vehicular communications channel modeling are reported in \cite{eckhardt2021channel}.

Research on THz communications should not wait until a perfect THz channel knowledge is established. Sufficiently accurate statistical channel models are crucial tools for researchers to address THz challenges and demonstrate practical solutions at an early stage. In fact, even after converging on accurate THz channel models, statistical models will remain attractive for conducting scalable simulations. Towards this end, we propose TeraMIMO, a statistical wideband UM-MIMO channel model that captures to a reasonable degree of accuracy most of the THz channel features. We believe that our simulator will advance signal processing for THz communication and sensing research, which remains relatively uncharted territory. When adopted by researchers from different backgrounds, TeraMIMO can establish a link between novel THz devices, THz channel and noise models, and THz signal processing techniques.

A number of mmWave, sub-THz, and THz system simulators are already publicly available. Most notably, NYUSIM \cite{Shihao2019} is a mmWave and sub-THz channel simulator that provides insight into important channel modeling components such as human blockage, outdoor-to-indoor penetration loss, and spatial consistency. Being measurement-based, the NYUSIM simulator has recently been extended to support a maximum operating frequency of $\unit[150]{GHz}$ and bandwidth of $\unit[800]{MHz}$. TeraSim \cite{hossain2018terasim} is another simulator of THz networks covering nanoscale and macroscale scenarios; it is a valuable ns-3 extension that implements physical and medium access THz control layer solutions. However, TeraSim does not capture many THz channel characteristics, and it does not support an end-to-end channel model. In particular, TeraSim does not account for  NLoS scenarios, antenna arrays, beamsteering and beamforming, spherical wave propagation, misalignment, and TV channel responses.

TeraMIMO significantly differs from existing simulators because it is mainly dealing with the physical link layer, and it supports the $3$D geometry of signal propagation.
However, it can be extended to support system-level simulations. Our simulator can thus complement TeraSim when implementing physical layer waveforms, for example.
In particular, TeraMIMO bears the following features:
\begin{enumerate}
    \item Provides a 3D end-to-end THz channel model which includes THz-specific peculiarities such as misalignment, spherical wave propagation, phase uncertainties in phase shifters, and beam split.
    \item Generates three propagation scenarios for LoS, NLoS, and LoS-dominant and NLoS-assisted communications. 
    \item Provides the required channel statistics such as coherence time, coherence bandwidth, maximum Doppler shift, and root mean square (RMS) delay spread.
    \item Accounts for frequency-selective (FS) THz channels over multiple bands and communication distances: From nano communications to short-range indoor/outdoor scenarios all the way to LoS links of hundred of meters.
    \item Adopts the array-of-subarrays (AoSA) antenna structure for hybrid beamforming and accounts for spatial sparsity.
    \item Supports three models of molecular absorption loss: An exact radiative transfer theory model for $\unit[0.1-10]{THz}$, and two approximations valid up to $\unit[450]{GHz}$.
    \item Provides different channel domain implementations, such as the delay domain and the frequency domain for the time-invariant (TIV) channel and the time-delay domain and time-frequency domain for the TV channel.
    \item Supports both an approximate planar wave model (PWM) and an accurate spherical wave model (SWM) that accurately accounts for the curvature of transmitted wavefronts in the near field.
    \item Introduces an efficient graphical user interface (GUI) for simulating multiple channel profiles.
\end{enumerate}

The remainder of this paper is organized as follows: We detail the system model in Sec.~\ref{sec:system_model}. Section~\ref{sec:channel_model} specifies the proposed THz channel model. We illustrate how to use the simulator and present sample simulation results in Sec.~\ref{sec:simulations}, before drawing conclusions and discussing future extensions in Sec.~\ref{sec:conclusions}. Concerning notation, bold upper case, bold lower case, and lower case letters correspond to matrices, vectors, and scalars, respectively. $\mbf{0}_{\mathrm{N}\times1}$ is a zero vector of size $\mathrm{N}\times \mathrm{1}$, $\mbf{I}_N$ is the identity matrix of size N. The superscripts
${(\cdot)}^\ast, {(\cdot)}^\mathrm{-1}$, ${(\cdot)}^\mathrm{T}$, and ${(\cdot)}^\mathcal{H}$ stand for the conjugate, inverse, transpose, and conjugate transpose functions, respectively. While, diag($\cdot$) denotes a block diagonal matrix. We use $\EE[\cdot]$ to denote the expectation, $\abs\cdot$ the absolute value, $\norm\cdot$ the Euclidean norm, $\norm{\cdot}_F$ the Frobenius norm, and $\PP(\cdot)$ the probability density function of a random variable. $\mathcal{CN}(\varpi,\sigma^2)$ is a complex Gaussian random variable with mean $\varpi$ and variance $\sigma^2$, and $j = \sqrt{-1}$ is the imaginary unit.
The superscripts ${(\mathrm{t})}$ and ${(\mathrm{r})}$ denote the transmitter (Tx) and receiver (Rx) parameters. We use $(\tilde{\cdot})$ and $(\cdot)$ to represent global and local coordinates, respectively, of position and AoD/AoA.
The variable notations are summarized in Tables~\ref{table:NotationsTableLatin} \ref{table:NotationsTableLatin2}, \ref{table:NotationsTableGreek}, and \ref{table:NotationsTableGreek2}.

\section{System Model}
\label{sec:system_model}

We propose a robust stochastic THz channel model that captures recent findings on THz propagation features and system architectures. We start by detailing the system model.

\subsection{Arrays of Subarrays}
\label{sec:aosa_model}

We consider a doubly massive MIMO system. This is a reasonable assumption for THz UM-MIMO because massive antenna arrays of small footprints can be deployed at both the Tx and Rx. THz UM-MIMO systems typically enable quasi-optical point-to-point (P2P) wireless aerial or terrestrial links in the backhaul or fronthaul of networks. We assume an AoSA architecture in which each subarray (SA) is formed of many antenna elements (AEs)~\cite{sarieddeen2021overview}. AoSA configurations can mitigate the high-frequency hardware constraints and support low-complexity beamforming; they combat the limited communication distance problem while maintaining a good spatial multiplexing gain \cite{LinAoSa}.
We further assume an AoSA structure to realize hybrid beamforming using a sub-connected structure in which analog beamforming is only conducted at the AEs in each SA. Each SA is fed by a single RF chain, reducing complexity and power consumption.

\begin{figure}[ht]
\centering
\includegraphics[width = 0.95\linewidth]{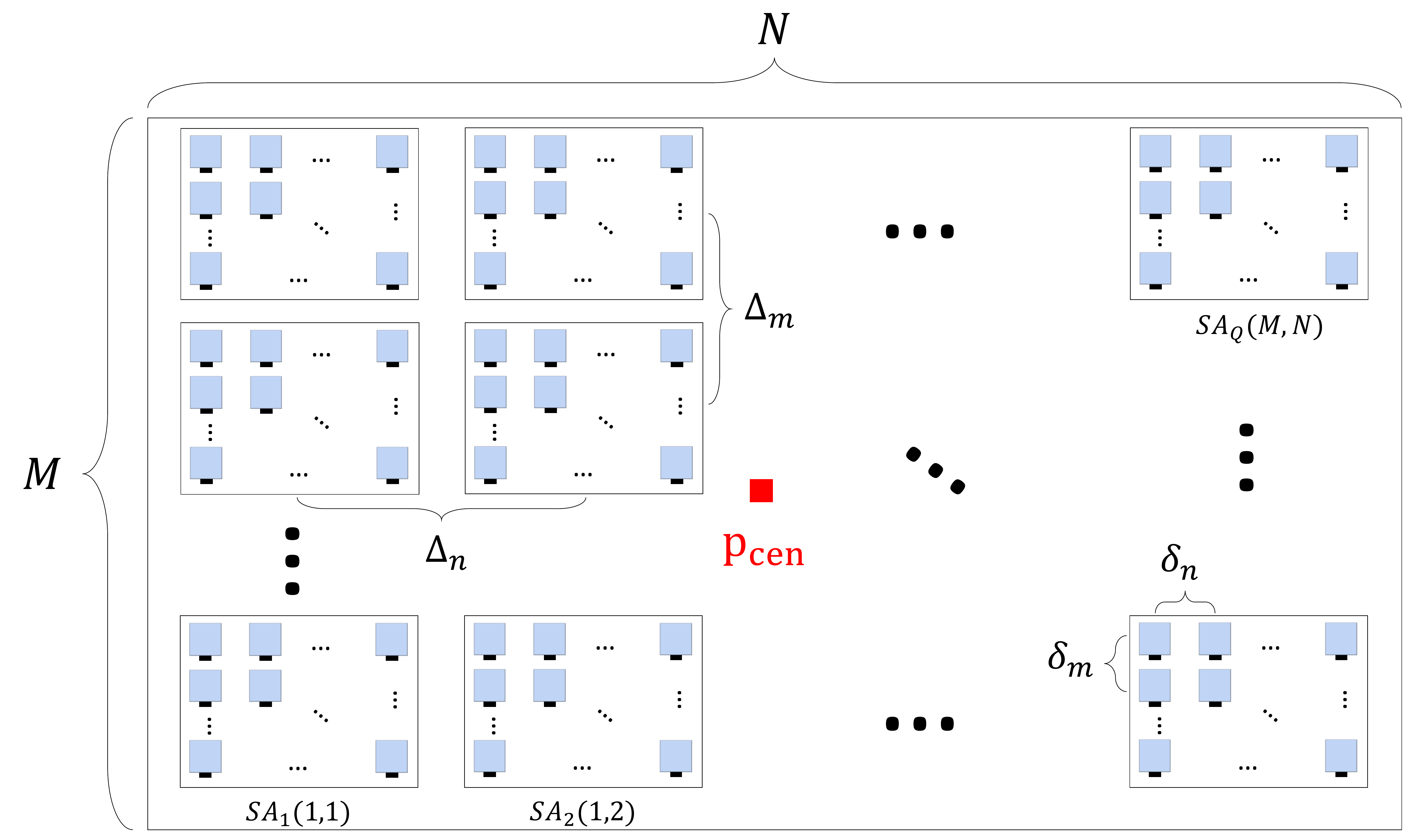}
\caption{Illustration of the AoSA structure.}
\label{fig:aosa_structure}
\end{figure}

We focus on planar AoSA structures consisting of $Q\!=\!M\!\times\!N$ SAs as shown in Fig.~\ref{fig:aosa_structure}. We define the index, $q$, of the SA at row $m$ ($1 \!\le\! m \!\le\! M$) and column $n$ ($1 \!\le\! n \!\le\! N$) via the map 
\begin{equation}
\label{eq:mapping_qmn}
q = (m-1)N+n, \ \ \ 1 \le q \le Q.
\end{equation}
In turn, the SA consists of $\bar{Q}\!=\!\bar{M}\times\bar{N}$ tightly-packed directional AEs. Each AE is attached to a wideband THz analog phase-shifter (PS) (implemented using graphene transmission lines in plasmonic solutions~\cite{lin2015indoor}).
We similarly define the index, $\bar{q}$, of the AE at row $\bar{m}$ ($1 \!\le\! \bar{m} \!\le\! \bar{M}$) and column $\bar{n}$ ($1 \!\le\! \bar{n} \!\le\! \bar{N}$) as
\begin{equation}
\label{eq:mapping_qmn2}
\bar{q} = (\bar{m}-1)\bar{N}+\bar{n}, \ \ \ 1 \le \bar{q} \le \bar{Q}.
\end{equation}
We denote by $\Delta_m$/$\Delta_n$ and $\delta_m$/$\delta_n$ the distances between the row/column centers of two adjacent SAs and AEs, respectively.

\subsection{Equivalent Baseband System Model}
\label{sec:baseband_model}

We consider a MC system, cyclic-prefix orthogonal frequency-division multiplexing (CP-OFDM) or an alternative MC scheme of non-overlapping subcarriers. Each of the $K$ subcarriers carries one data symbol per SA, forming a length-$K$ data symbol stream. The total number of transmitted streams is $N_{\mathrm{S}}$, where $N_{\mathrm{S}}\leq Q$ is the total number of RF chains. The information-bearing symbol vector per subcarrier $k\!=\!\{0,\cdots,K-1\}$ is $\mbf{s}[k]\!=\!{\left[{s_1,s_2,\cdots,s_{N_{\mathrm{S}}}}\right]}^\Tr \!\in\! \mathcal{X}^{N_{\mathrm{S}}\times1}$, which consists of data symbols drawn from a quadrature amplitude modulation (QAM) constellation, $\mathcal{X}$. We assume normalized symbols such that $\EE[\mbf{s}[k]\mbf{s}^\ast[k]]\!=\!\frac{P_{\mathrm{S}}}{KN_{\mathrm{S}}}\mbf{I}_{N_{\mathrm{S}}}$, where $P_{\mathrm{S}}$ is the total average transmit power over all subcarriers. 

Digital baseband beamforming defines the utilization of SAs, routing data streams via RF chains. In highly correlated P2P THz links, efficient nonlinear techniques such as block multi-diagonalization \cite{Nakagawa2017}, or THz-specific FS hybrid beamforming \cite{Yuan2020} are required. Hence, $\mbf{s}[k]$ is first precoded using $\mbf{W}_{\BBand}[k]\!\in\!\CC^{Q^{(\mathrm{t})}\times N_{\mathrm{S}}}$, followed by analog RF beamforming using $\mbf{W}_{\RF}\!\in\!\CC^{Q^{(\mathrm{t})}\bar{Q}^{(\mathrm{t})}\times Q^{(\mathrm{t})}}$. Spatial energy focusing is achieved via analog beamforming over many AEs per SA, where beamsteering codebook designs can be utilized per RF chain (digitally controlled THz PSs). We have $\mbf{W}_{\RF}\!=\!\mathrm{diag}({\mbf{w}_1, \mbf{w}_2,\cdots,\mbf{w}_{Q^{(\mathrm{t})}}})$, and $\mbf{w}_q^{(\mathrm{t})}$ is a $\bar{Q}^{(\mathrm{t})}\!\times\!1$ vector
(detailed in Sec. \ref{sec:sa_ear_sv_bm}). Assuming frequency-flat analog beamforming, the discrete-time transmitted complex baseband signal at the $\nth{k}$ subcarrier is expressed as
\begin{equation}
    \mbf{x}[k]=\mbf{W}_{\RF}\mbf{W}_{\BBand}[k]\mbf{s}[k].
    \label{eq:tx_subc_sig}
\end{equation}
Denoting by $\mbf{H}[k]\!\in\!\CC^{Q^{(\mathrm{r})}\bar{Q}^{(\mathrm{r})}\times Q^{(\mathrm{t})}\bar{Q}^{(\mathrm{t})}}$ the overall complex channel matrix at the $\nth{k}$ subcarrier, and assuming a FS-TIV scenario, the overall UM-MIMO channel matrix can be expressed as 
\begin{equation}
    \mbf{H}[k]= \begin{bmatrix}
    \mbf{H}_{1,1}[k]& \cdots & \mbf{H}_{1,Q^{(\mathrm{t})}}[k]\\
    \vdots & \ddots & \vdots\\
     \mbf{H}_{Q^{(\mathrm{r})},1}[k] & \cdots & \mbf{H}_{Q^{(\mathrm{r})},Q^{(\mathrm{t})}}[k]\\
    \end{bmatrix},
    \label{eq:overall_H_sub_H}
\end{equation}
where $\mbf{H}_{q^{(\mathrm{r})},q^{(\mathrm{t})}}[k]\!\in\!\CC^{\bar{Q}^{(\mathrm{r})}\times\bar{Q}^{(\mathrm{t})}}$ is the channel sub-matrix between the $\nth{q^{(\mathrm{t})}}$ Tx SA and the $\nth{q^{(\mathrm{r})}}$ Rx SA. The $\nth{k}$-subcarrier received signal can then be expressed as
\begin{equation}
    \mbf{y}[k]=\mbf{C}^\Hr_{\BBand}[k]\mbf{C}^\Tr_{\RF}\mbf{H}[k]\mbf{x}[k]+\mbf{C}^\Hr_{\BBand}[k]\mbf{C}^\Tr_{\RF}\mbf{n}[k].
    \label{eq:rx_subc_sig}
\end{equation}

Here, assuming perfect synchronization, the received signal is processed using an RF combining matrix $\mbf{C}_{\RF}\!\in\!\CC^{Q^{(\mathrm{r})}\bar{Q}^{(\mathrm{r})}\times Q^{(\mathrm{r})}}$ and a digital baseband combining matrix $\mbf{C}_{\BBand}[k]\!\in\!\CC^{Q^{(\mathrm{r})}\times N_{\mathrm{S}}}$ (same constraints on combiners as beamformers).
We denote by $\mbf{n}[k]\!\in\!\CC^{Q^{(\mathrm{r})}\bar{Q}^{(\mathrm{r})}\times 1}$ the additive white Gaussian noise (AWGN) vector of independently distributed $\mathcal{CN}\left(\mbf{0},\sigma_n^2\mbf{I}_{Q^{(\mathrm{r})}\bar{Q}^{(\mathrm{r})}}\right)$ elements, where $\sigma_n^2$ is the noise power.

\subsection{Coordinate System}
\label{sec:global_local_system}

\begin{figure}[t]
\centering
\includegraphics[width = 0.928\linewidth]{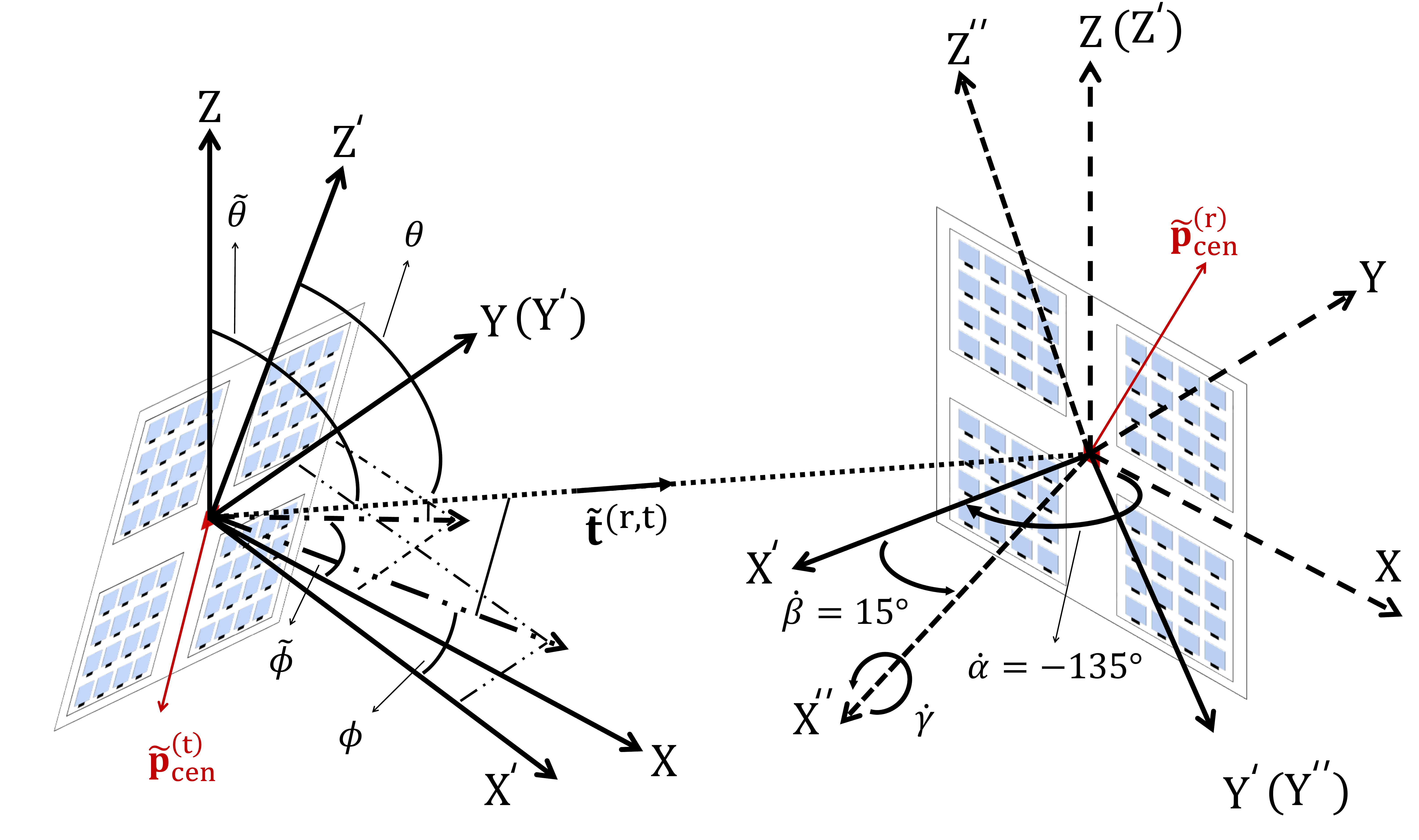}
\caption{An illustration of global/local coordinate systems.} 
\label{fig:aosa_tx_rx}
\end{figure}

A practical THz communication system consists of several UM-MIMO base-stations (BSs) and user-equipment (UEs) operating at the same time and frequency resources. Hence, it is essential to define an accurate global coordinate system; we assume a right-handed 3D Cartesian coordinate system $\mathrm{XYZ}$, and we set the arrays to lie on the Y-Z plane in the local coordinate, as shown in Fig.~\ref{fig:aosa_tx_rx}. We designate the centers of the Tx and Rx to be located in the global Cartesian coordinate systems at $\tilde{\mbf{p}}_{\cent}^{(\mathrm{t})} \!=\! [\tilde{p}_{x}^{(\mathrm{t})},\tilde{p}_{y}^{(\mathrm{t})}, \tilde{p}_{z}^{(\mathrm{t})}]^\Tr$ and $\tilde{\mbf{p}}_{\cent}^{(\mathrm{r})}\! =\! [\tilde{p}_{x}^{(\mathrm{r})},\tilde{p}_{y}^{(\mathrm{r})},\tilde{p}_{z}^{(\mathrm{r})}]^\Tr$, and the centers of the AoSA to be the origins of the local Cartesian coordinate systems (see Fig.~\ref{fig:aosa_tx_rx}).

We use Euler angles to describe the local coordinate system's orientation with respect to the global coordinate system. We adopt the Tait-Bryan angles with intrinsic right-handed rotations of the order $\mathrm{ZYX}$.
Intrinsic rotations are about the local axes, and extrinsic rotations are about the global axes; an intrinsic rotation sequence is an inversed extrinsic rotation sequence.
The angles $\dot{\alpha}\!\in\!(-\pi,\pi]$, $\dot{\beta}\!\in\![-\pi/2,\pi/2]$, and $\dot{\gamma}\!\in\!(-\pi,\pi]$ represent rotations around the $\mathrm{Z}$-, $\mathrm{Y}$-, and $\mathrm{X}$-axis of the local coordinate system, respectively. A sample rotation is illustrated in Fig.~\ref{fig:aosa_tx_rx}: For the receiver, the AoSA first rotates $\dot{\alpha}\!=\!-135^\circ$ about the $\mathrm{Z}$-axis, then $\dot{\beta}\!=\!15^\circ$ and $\dot{\gamma}$ about the $\mathrm{Y}$- and $\mathrm{X}$"-axis, respectively (positive/negative signs indicate counterclockwise/clockwise rotations). The corresponding rotation matrix can be represented as
\begin{equation}
    \label{eq:eq_rot_mat}
    \mbf{R} = 
    \begin{bmatrix}
    c_{\dot{\alpha}} c_{\dot{\beta}} & c_{\dot{\alpha}} s_{\dot{\beta}} s_{\dot{\gamma}}-c_{\dot{\gamma}} s_{\dot{\alpha}} & s_{\dot{\alpha}} s_{\dot{\gamma}} +c_{\dot{\alpha}} c_{\dot{\gamma}} s_{\dot{\beta}}\\
    c_{\dot{\beta}} s_{\dot{\alpha}}  & c_{\dot{\alpha}} c_{\dot{\gamma}} +s_{\dot{\alpha}} s_{\dot{\beta}} s_{\dot{\gamma}}  & c_{\dot{\gamma}} s_{\dot{\alpha}} s_{\dot{\beta}} -c_{\dot{\alpha}} s_{\dot{\gamma}}\\
    -s_{\dot{\beta}} & c_{\dot{\beta}} s_{\dot{\gamma}}  & c_{\dot{\beta}} c_{\dot{\gamma}} 
    \end{bmatrix},
\end{equation}
where $c_{\dot{\alpha}}$ and $s_{\dot{\alpha}}$ are short for $\cos(\dot{\alpha})$ and $\sin(\dot{\alpha})$, respectively.
Note that $\mbf{R}$ is real and orthogonal ($\tranb{R}=\invb{R}$).
The location of the $\nth{q}$ SA in local coordinates is
\begin{equation}
    {\mbf{p}}_{q} =
    \begin{bmatrix}
    0\\
    (n-1-\frac{N-1}{2})\Delta_{n}\\
    (m-1-\frac{M-1}{2})\Delta_{m}
    \end{bmatrix},
    \label{eq:local_sa_position}
\end{equation}
and the location of the $\nth{\bar{q}}$ AE of the $\nth{q}$ SA is given by
\begin{equation}
    {\mbf{p}}_{q,\bar{q}} ={\mbf{p}}_{q} +
    \begin{bmatrix}
    0\\
    (\bar{n}-1-\frac{\bar{N}-1}{2})\delta_{n}\\
    (\bar{m}-1-\frac{\bar{M}-1}{2})\delta_{m}
    \end{bmatrix}.
    \label{eq:local_ae_position}
\end{equation}
The global positions of an SA/AE denoted by ${\tilde{\mbf{p}}}_{q}/{\tilde{\mbf{p}}}_{q,\bar{q}}$ are related to the local positions through the rotation matrix $\mbf{R}$ by
\begin{equation}
    \label{eq:eq_sa_glob}
    \tilde{\mbf{p}}_{q}=\mbf{R}{\mbf{p}}_{q}+\tilde{\mbf{p}}_{\cent}, \ \ \ \ \tilde{\mbf{p}}_{q,\bar{q}} = \mbf{R}{\mbf{p}}_{q,\bar q}+\tilde{\mbf{p}}_{\cent}.
\end{equation}
Conversely, the local positions can be retrieved from the global positions using the relationship
\begin{equation}
    \label{eq:eq_sa_local_using_global}
    {\mbf{p}}_{q}=\invb{R}(\tilde{\mbf{p}}_{q}-\tilde{\mbf{p}}_{\cent}), \ \ \ \ {\mbf{p}}_{q,\bar{q}}=\invb{R}(\tilde{\mbf{p}}_{q,\bar q}-\tilde{\mbf{p}}_{\cent}).
\end{equation}
If the distance between the Tx and Rx AoSA is much larger than the array size, i.e., $\norm{\tilde{\mbf{p}}_{\cent}^{(\mathrm{r})}-\tilde{\mbf{p}}_{\cent}^{(\mathrm{t})}}\gg \max(M\Delta_m, N\Delta_n)$, the global AoD direction $\tilde{\mbf{t}}^{(\mathrm{r},\mathrm{t})}$ and the local AoD direction ${\mbf{t}}^{(\mathrm{r},\mathrm{t})}$ from each SA can be represented as the unit direction vectors
\begin{align}
    \tilde{\mbf{t}}^{(\mathrm{r},\mathrm{t})} &=
    \begin{bmatrix}
    \tilde{{t}}_1\\
    \tilde{{t}}_2\\
    \tilde{{t}}_3
    \end{bmatrix}
    = \frac{\tilde{\mbf{p}}_{\cent}^{(\mathrm{r})}-\tilde{\mbf{p}}_{\cent}^{(\mathrm{t})}}{\norm{\tilde{\mbf{p}}_{\cent}^{(\mathrm{r})}-\tilde{\mbf{p}}_{\cent}^{(\mathrm{t})}}},\\
    {\mbf{t}^{(\mathrm{r},\mathrm{t})}} &=\invb{R}\tilde{\mbf{t}}^{(\mathrm{r},\mathrm{t})}.
    \label{eq:dir_vec_from_position}
\end{align}
Similarly, the global and local AoA direction vectors can be obtained from $\tilde{\mbf{t}}^{(\mathrm{t},\mathrm{r})}\!=\!-\tilde{\mbf{t}}^{(\mathrm{r},\mathrm{t})}$ and ${\mbf{t}^{(\mathrm{t},\mathrm{r})}} \!=\!\invb{R}\tilde{\mbf{t}}^{(\mathrm{t},\mathrm{r})}$, respectively.
We adopt the local coordinate systems to represent AoD/AoA in the remainder of the paper, for convenience.
The AoD/AoA vectors are defined using the azimuth and elevation angles. The azimuth angle, $\phi\!\in\![-\pi,\pi)$, is the angle between the projection of $\mbf{t}$ on the X-Y plane and the X-axis. The elevation angle, $\theta\!\in\![0,\pi]$, is the angle between $\mbf{t}$ and the Z-axis, as shown in Fig.~\ref{fig:aosa_tx_rx}.
Therefore, $\mbf{t}$ can be expressed as
\begin{equation}
    \mbf{t} = \begin{bmatrix}
    {t}_1\\
    {t}_2\\
    {t}_3
    \end{bmatrix} = 
    \begin{bmatrix}
    \cos(\phi)\sin(\theta)\\
    \sin(\phi)\sin(\theta)\\
    \cos(\theta)
    \end{bmatrix},
    \label{eq:eq_local_dir_angle}
\end{equation}
and the angles can be extracted from $\mbf{t}$ as
\begin{equation}
    \begin{bmatrix}
        \phi\\
        \theta
    \end{bmatrix}=
    \begin{bmatrix}
        \atantwo(t_2,t_1)\\
        \arccos(t_3)
    \end{bmatrix},
    \label{eq:eq_local_angle}
\end{equation}
where $\atantwo(\cdot)$ is the four-quadrant inverse tangent.

\section{Channel Model}
\label{sec:channel_model}

This section details the proposed stochastic THz ultra-wideband UM-MIMO channel model. We capture the LoS path deterministically and generate the MP components using random processes of known distributions, thus combining the advantages of capturing the environment's geometry in RT~\cite{Han2015MR} and maintaining the simulator's flexibility in statistical modeling~\cite{priebe2013stochasticchannelmodel,lin2015indoor}. We adopt a statistical tap-delay profile suitable for ultra-broadband channels to model the impulse response and MP parameters \cite{Han_channelmodels}.

\subsection{THz Channel Characteristics}
\label{sec:tivch_domain}

Channel modeling involves characterizing both large- and small-scale fading. Large-scale fading includes blockage and shadowing in system-level simulations. However, accurately modeling small-scale fading and the path and molecular absorption losses is sufficient for simulating link-level channels.
Fading can be selective or flat in the frequency domain, fast or slow in the time domain, which requires clearly defining the time spread of signals and the channel's time variance. We define the channel in the delay (time-delay) domain, $\mbf{H}(\tau)$, and the frequency domain, $\mbf{H}(f)$, with both domains being related via a Fourier transform. We further define the channel in the time domain, $\mbf{H}(t)$, and the Doppler (Doppler-shift) domain, $\mbf{H}(\nu)$, which are also interchangeable via a Fourier transform. Note that the time- and frequency-domain channels are duals \cite{sklar1997rayleigh}, and moving between the delay-Doppler and time-frequency channel representations can be achieved by using the symplectic finite Fourier transform \cite{hadani2017orthogonal}.

The physical layer design depends on channel parameters in both time/delay and Doppler/frequency domains. The coherence bandwidth, $B_{\coh}\!=\!\frac{1}{5\tau_{\rms}}$, is the maximum frequency span over which the channel correlation exceeds 0.5, where $\tau_{\rms}$ is the RMS delay spread. Flat-fading occurs when the Tx symbol duration is much larger than the delay spread ($T_{\sym}\!\gg\!\tau_{\rms}$) or when the coherence bandwidth is much larger than the signal bandwidth ($B_{\coh}\!\gg\! B_{\sig}$). Similarly, the coherence time, $T_{\coh}\!=\!\sqrt{\frac{9}{16\pi f_{\maxx}}\times\frac{1}{f_{\maxx}}}$, is the maximum time span over which the fading gain magnitude does not change significantly, where $f_{\maxx}\!=\!\frac{\vartheta}{c_0}f_c$ is the maximum Doppler shift, $\vartheta$ is the velocity of the UE or BS, in $\unit{(m/sec)}$, $c_0$ is the speed of light, and $f_c$ is the carrier frequency. The symbol duration is typically designed to be much less than the coherence time ($T_{\sym}\!\ll\! T_{\coh}$). 

In favorable near-static THz environments, we expect TIV and flat-fading channels. The delay spread is small with high antenna directivity, which increases the coherence bandwidth and the likelihood of flat fading. However, the channel can still be both TV and FS in non-static THz environments. MP components and molecular absorption (in wideband systems) cause THz frequency selectivity. In medium-distance indoor THz environments, residual MP components persist, limiting $B_{\coh}$ to $\unit[1-5]{GHz}$~\cite{Han2015MR,Han2016M-W-D-A,Han2016_DABARA}, whereas the width of absorption-free bands can exceed $100$ Gigahertz ($\GHz$). Furthermore, a mobility as slow as $\unit[1]{m/sec}$ induces a Doppler shift of approximately $1$ Kilohertz (KHz) at $\unit[0.3]{THz}$, which exceeds $T_{\coh}\!=\!\unit[0.42]{ms}$.

We construct our FS THz channel model over multiple subcarriers, each of which is divided into multiple sub-bands, with LoS and NLoS components per subcarrier. The delay-$u$ TIV channel matrix between the Tx and Rx SAs, $q^{(\mathrm{t})}$ and $q^{(\mathrm{r})}$, for a communication distance, $d$, and subcarrier center frequency, $f_k$, can thus be expressed as~\cite{Han2018UMMIMO,lin2015indoor,Yuan2020}
\begin{equation}
\begin{aligned}
    \label{eq:ch_delay_domain}
    \mbf{H}_{q^{(\mathrm{r})},q^{(\mathrm{t})}}^u(f_k,d)&=\alpha^{\LoS}(f_k,d)G^{(\mathrm{t})}(\mbf{\Phi}^{(\mathrm{t})})G^{(\mathrm{r})}(\mbf{\Phi}^{(\mathrm{r})})\\
    &\times\mbf{a}^{(\mathrm{r})}(\mbf{\Phi}^{(\mathrm{r})})\mbf{a}^{(\mathrm{t})^{\Tr}}(\mbf{\Phi}^{(\mathrm{t})})\mathrm{dirac}(uT_s-\tau^{\LoS})\\
    &+\sum_{c=1}^{N_{\clu}}\sum_{\ell=1}^{N_{\ray}^{c}}\alpha_{c,\ell}^{\NLoS}(f_k,d)G^{(\mathrm{t})}(\mbf{\Phi}_{c,\ell}^{(\mathrm{t})}) G^{(\mathrm{r})}(\mbf{\Phi}_{c,\ell}^{(\mathrm{r})})\\
    &\times\mbf{a}^{(\mathrm{r})}(\mbf{\Phi}_{c,\ell}^{(\mathrm{r})})\mbf{a}^{(\mathrm{t})^{\Tr}}(\mbf{\Phi}_{c,\ell}^{(\mathrm{t})})\mathrm{dirac}(uT_s-\tau^{\NLoS}_{c,\ell}),
\end{aligned}
\end{equation}
where $\alpha^{\LoS}$ and $\alpha_{c,\ell}^{\NLoS}$ denote the path gains of the LoS and NLoS rays, $G^{(\mathrm{t})}(\cdot)$ and $G^{(\mathrm{r})}(\cdot)$ are the Tx and Rx antenna gains, and $\mbf{a}^{(\mathrm{t})}(\cdot)\in\!\CC^{\bar{Q}^{(\mathrm{t})}\times 1}$ and $\mbf{a}^{(\mathrm{r})}(\cdot)\in\!\CC^{\bar{Q}^{(\mathrm{r})}\times 1}$ are the Tx and Rx antenna array response vectors, respectively (called beamsteering vectors in this work).
\begin{figure*}[ht]
\begin{equation}
\begin{aligned}
    \label{eq:ch_freq_domain}
    \mbf{H}_{q^{(\mathrm{r})},q^{(\mathrm{t})}}[k]&=\sum_{u=0}^{U-1}\mbf{H}_{q^{(\mathrm{r})},q^{(\mathrm{t})}}^u(f_k,d)e^{-j\frac{2\pi k}{K}u} =\alpha^{\LoS}G^{(\mathrm{t})}({\mbf{\Phi}}^{(\mathrm{t})})G^{(\mathrm{r})}({\mbf{\Phi}}^{(\mathrm{r})})\mbf{a}^{(\mathrm{r})}(\mbf{\Phi}^{(\mathrm{r})})\mbf{a}^{(\mathrm{t})^{\Tr}}(\mbf{\Phi}^{(\mathrm{t})})\eta(k,\tau^{\LoS})
    \\&+\sum_{c=1}^{N_{\clu}}\sum_{\ell=1}^{N_{\ray}^{c}}\alpha_{c,\ell}^{\NLoS}G^{(\mathrm{t})}(\mbf{\Phi}_{c,\ell}^{(\mathrm{t})})G^{(\mathrm{r})}(\mbf{\Phi}_{c,\ell}^{(\mathrm{r})})\mbf{a}^{(\mathrm{r})}(\mbf{\Phi}_{c,\ell}^{(\mathrm{r})})\mbf{a}^{(\mathrm{t})^{\Tr}}(\mbf{\Phi}_{c,\ell}^{(\mathrm{t})})\eta(k,\tau^{\NLoS}_{c,\ell})
    \\&=\alpha^{\LoS}G^{(\mathrm{t})}({\mbf{\Phi}}^{(\mathrm{t})})G^{(\mathrm{r})}({\mbf{\Phi}}^{(\mathrm{r})})\mbf{a}^{(\mathrm{r})}(\mbf{\Phi}^{(\mathrm{r})})\mbf{a}^{(\mathrm{t})^{\Tr}}(\mbf{\Phi}^{(\mathrm{t})})e^{-j 2\pi \frac{kB}{K}\tau^{\LoS}}
    \\&+\sum_{c=1}^{N_{\clu}}\sum_{\ell=1}^{N_{\ray}^{c}}\alpha_{c,\ell}^{\NLoS}G^{(\mathrm{t})}(\mbf{\Phi}_{c,\ell}^{(\mathrm{t})})G^{(\mathrm{r})}(\mbf{\Phi}_{c,\ell}^{(\mathrm{r})})\mbf{a}^{(\mathrm{r})}(\mbf{\Phi}_{c,\ell}^{(\mathrm{r})})\mbf{a}^{(\mathrm{t})^{\Tr}}(\mbf{\Phi}_{c,\ell}^{(\mathrm{t})})e^{-j 2\pi \frac{kB}{K}\tau^{\NLoS}_{c,\ell}}.
\end{aligned}
\end{equation}
\end{figure*}
We denote by $N_{\clu}$ and $N_{\ray}^{c}$ the number of clusters and rays in the $\nth{c}$ cluster, by $\mbf{\Phi}^{(\mathrm{t})}\!=\![\phi^{(\mathrm{t})},\theta^{(\mathrm{t})}]^\Tr$/$\mbf{\Phi}^{(\mathrm{r})}\!=\![\phi^{(\mathrm{r})},\theta^{(\mathrm{r})}]^\Tr$ and $\mbf{\Phi}_{c,\ell}^{(\mathrm{t})}\!=\![\phi_{c,\ell}^{(\mathrm{t})},\theta_{c,\ell}^{(\mathrm{t})}]^\Tr$/$\mbf{\Phi}_{c,\ell}^{(\mathrm{r})}\!=\![\phi_{c,\ell}^{(\mathrm{r})},\theta_{c,\ell}^{(\mathrm{r})}]^\Tr$ the LoS and $\nth{\ell}$ ray $\nth{c}$ cluster NLoS angle of departure/arrival vectors, and by $\! \tau^{\LoS} \!=\! {d_{q^{(\mathrm{r})},q^{(\mathrm{t})}}/c_{0}}$ and $\tau^{\NLoS}_{c,\ell}$ the corresponding ToA of the LoS and NLoS rays, respectively (see Sec.~\ref{sec:sa_ear_sv_bm} for more details). We denote by $\mathrm{dirac}(\tau)$ the dirac function for $T_s$-spaced signaling evaluated at $\tau$ seconds.
The $\nth{k}$ subcarrier frequency is expressed as $f_k\!=\!f_{c}\!+\!\frac{B}{K}(k\!-\!\frac{K-1}{2})$, where $f_{c}$ is the center frequency and $B$ is the total bandwidth (divided into $K$ subcarriers).
Furthermore, to obtain high frequency resolution in the FS channel, we compute the response over $N_{\subb}$ sub-bands within each subcarrier, where $f_{n_{\subb}}\!=\!f_{k}\!+\!\frac{B_{\subb}}{N_{\subb}}(n_{\subb}\!-\!\frac{N_{\subb}-1}{2})$, and $B_{\subb}\!=\!\frac{B}{K}$ is the sub-band bandwidth. The frequency-domain channel at the $\nth{k}$ subcarrier can be obtained form the delay-$u$ channel \eqref{eq:ch_delay_domain} via a Fourier transformation as illustrated in~\eqref{eq:ch_freq_domain}, where $\eta(k,\tau)\!=\!\sum_{u=0}^{U-1}\mathrm{dirac}(uT_s-\tau)e^{-j\frac{2\pi k}{K}u}$. Note that Eqs. \eqref{eq:ch_delay_domain} and \eqref{eq:ch_freq_domain} can be augmented with a pulse-shaping filter for both Tx and Rx \cite{alkhateeb2016frequency}.  With a limited number of NLoS rays at THz frequencies ~\cite{Han2016_DABARA}, $\mbf{H}[k]$ is low-rank. The channel coefficients at different subcarriers are also highly correlated, where the SA steering vectors are fixed across all subcarriers~\cite{Yuan2020}.

For analog beamforming per SA, given the target AoD and AoA, $\mbf{\Phi}_{0}^{(\mathrm{t})}\!=\![\phi_{0}^{(\mathrm{t})},\theta_{0}^{(\mathrm{t})}]^\Tr$ and $\mbf{\Phi}_{0}^{(\mathrm{r})}\!=\![\phi_{0}^{(\mathrm{r})},\theta_{0}^{(\mathrm{r})}]^\Tr$, the corresponding beamforming vectors are denoted by $\hat{\mbf{a}}^{(\mathrm{t})}(\mbf{\Phi}_{0}^{(\mathrm{t})})$ and $\hat{\mbf{a}}^{(\mathrm{r})}(\mbf{\Phi}_{0}^{(\mathrm{r})})$ (to be defined in Sec.~\ref{sec:sa_ear_sv_bm}), respectively. Following analog beamforming, the effective baseband channel, $\hat{\mbf{H}}[k]$, is
\begin{equation}
  \begin{aligned}
    \label{eq:H_ch_sa_freq_domain}
    &\hat{\mbf{H}}[k]
    =\mbf{C}^{\Tr}_{\RF}\mbf{H}[k]\mbf{W}_{\RF}\\
    &=\!\mbf{C}^{\Tr}_{\RF}\mbf{H}[k]\!\begin{bmatrix}\hat{\mbf{a}}^{(\mathrm{t})}({\mbf{\Phi}_{0}}_{1,1}^{(\mathrm{t})})& \mbf{0}_{\bar{Q}^{(\mathrm{t})}\times1} & \cdots & \mbf{0}_{\bar{Q}^{(\mathrm{t})}\times1}\\ \!\mbf{0}_{\bar{Q}^{(\mathrm{t})}\times1} & \!\hat{\mbf{a}}^{(\mathrm{t})}({\mbf{\Phi}_{0}}_{2,2}^{(\mathrm{t})})&\!\vdots & \!\mbf{0}_{\bar{Q}^{(\mathrm{t})}\times1}\\ \!\!\!\!\!\!\vdots\!\!\! & \!\!\!\!\!\!\vdots\!\!\!& \!\!\!\!\!\!\ddots\!\!\! & \!\!\!\!\!\!\vdots\!\!\!\\ \!\mbf{0}_{\bar{Q}^{(\mathrm{t})}\times1}& \!\mbf{0}_{\bar{Q}^{(\mathrm{t})}\times1} & \!\cdots&\hat{\mbf{a}}^{(\mathrm{t})}({\mbf{\Phi}_{0}}_{Q^{(\mathrm{t})},Q^{(\mathrm{t})}}^{(\mathrm{t})})\\\end{bmatrix}\\
    &=\begin{bmatrix}\hat{h}_{1,1}[k]& \hat{h}_{1,2}[k] & \cdots & \hat{h}_{1,Q^{(\mathrm{t})}}[k]\\\vdots & \vdots& \ddots &\vdots\\\hat{h}_{Q^{(\mathrm{r})},1}[k]&\cdots & \cdots&\hat{h}_{Q^{(\mathrm{r})},Q^{(\mathrm{t})}}[k]\\\end{bmatrix},
  \end{aligned}
\end{equation}
where the equivalent frequency-domain baseband channel between the $\nth{q^{(\mathrm{t})}}$ Tx and $\nth{q^{(\mathrm{r})}}$ Rx SAs is expressed as
\begin{equation}
  \begin{aligned}
    \label{eq:ch_sa_freq_domain}
    \hat{h}_{q^{(\mathrm{r})},q^{(\mathrm{t})}}[k]&=\hat{\mbf{a}}^{(\mathrm{r})^{\Tr}}({\mbf{\Phi}_{0}}_{q^{(\mathrm{r})},q^{(\mathrm{r})}}^{(\mathrm{r})})\mbf{H}_{q^{(\mathrm{r})},q^{(\mathrm{t})}}[k]\hat{\mbf{a}}^{(\mathrm{t})}({\mbf{\Phi}_{0}}_{q^{(\mathrm{t})},q^{(\mathrm{t})}}^{(\mathrm{t})})\\
    &=\alpha^{\LoS}(k)G^{(\mathrm{t})}(\mbf{\Phi}^{(\mathrm{t})})G^{(\mathrm{r})}(\mbf{\Phi}^{(\mathrm{r})})\\
    &\times\mathcal{A}_{eq}^{(\mathrm{r})}(\mbf{\Phi}^{(\mathrm{r})},\mbf{\Phi}_{0}^{(\mathrm{r})})\mathcal{A}_{eq}^{(\mathrm{t})^{\Tr}}(\mbf{\Phi}^{(\mathrm{t})},\mbf{\Phi}_{0}^{(\mathrm{t})})e^{-j 2\pi \frac{kB}{K}\tau^{\LoS}}\\
    &+\sum_{c=1}^{N_{\clu}}\sum_{\ell=1}^{N_{\ray}^{c}}\alpha_{c,\ell}^{\NLoS}(k)G^{(\mathrm{t})}(\mbf{\Phi}_{c,\ell}^{(\mathrm{t})})G^{(\mathrm{r})}(\mbf{\Phi}_{c,\ell}^{(\mathrm{r})})\\
    &\times\mathcal{A}_{eq}^{(\mathrm{r})}(\mbf{\Phi}_{c,\ell}^{(\mathrm{r})},\mbf{\Phi}_{0}^{(\mathrm{r})})\mathcal{A}_{eq}^{(\mathrm{t})^{\Tr}}(\mbf{\Phi}_{c,\ell}^{(\mathrm{t})},\mbf{\Phi}_{0}^{(\mathrm{t})})e^{-j 2\pi \frac{kB}{K}\tau^{\NLoS}_{c,\ell}}.
  \end{aligned}
\end{equation}
$\mathcal{A}_{eq}^{\mathrm{(t)}}$ and $\mathcal{A}_{eq}^{\mathrm{(r)}}$ are the equivalent array responses (to be defined in Sec.~\ref{sec:sa_ear_sv_bm}). Note that we use a transpose instead of conjugate transpose in the beamsteering and beamforming vectors because of the definition of AoA and AoD in our system model. Spatial tuning techniques of antenna element separations have been recently investigated as a means to guarantee THz channel orthogonality and achieve high multiplexing gains \cite{sarieddeen2021terahertz,Sarieddeen2019SM}.

\subsection{Path Gains and Multipath Components}
\label{sec:path_gains}

In most RT-based THz channel models~\cite{han_channel_2016,Han2015MR,Moldovan2014}, the LoS and reflected rays are included, while the scattered and diffracted rays are assumed to add insignificant contributions to the received signal power \cite{Han2018UMMIMO}. The LoS path loss between the Tx and Rx SAs, $q^{(\mathrm{t})}$ and $q^{(\mathrm{r})}$, is expressed as
\begin{align}\label{eq:LoS_pg}
    \alpha^{\LoS}({q^{(\mathrm{r})},q^{(\mathrm{t})}})=\left(\frac{c_0}{4\pi f_{k} d_{q^{(\mathrm{r})},q^{(\mathrm{t})}}}\right)^{\frac{\gamma}{2}}e^{-\frac{1}{2}\mathcal{K}(f_{k})d_{q^{(\mathrm{r})},q^{(\mathrm{t})}}}, 
\end{align}
where $d_{q^{(\mathrm{r})},q^{(\mathrm{t})}}$ is the distance separating the SAs and $\mathcal{K}(f_{k})$ is the frequency-dependent molecular absorption coefficient. Here, the factor $\left(\frac{c_0}{4\pi f_k d_{q^{(\mathrm{r})},q^{(\mathrm{t})}}}\right)^{\frac{\gamma}{2}}$ accounts for the spreading loss, while $e^{-\frac{1}{2}\mathcal{K}(f_{k})d_{q^{(\mathrm{r})},q^{(\mathrm{t})}}}$ accounts for the molecular absorption loss. The path loss exponent, $\gamma$, is usually set to 2 in free space. In many measurement-based mmWave and sub-THz works, its best fit value is also around 2 ($2.2$ in \cite{Abbasi2020}, for example). This is different from low-frequency regimes where $\gamma$ heavily depends on the environment, with typical values of 3 and 4.5 for flat rural and dense urban environments, respectively. The NLoS path gain is computed as
\begin{align}\label{eq:NLoS_pg}
    \alpha_{c,\ell}^{\NLoS}(f_k,d) = \abs{\alpha_{c,\ell}^{\NLoS}(f_k,d)}e^{j\beta_{c,\ell}},
\end{align} 
where $\beta_{c,\ell}$ is a phase shift uniformly distributed over $[0,2 \pi)$.

The NLoS components are crucial for enabling THz connectivity when the LoS path is obstructed, which is highly probable with narrow THz beamwidths. The mmWave, sub-THz, and THz channels are assumed to retain a few NLoS paths due to high reflection losses and beamforming gains. However, THz channels are much sparser in the angular domain, with a much smaller overall angular spread. The number of MP THz components is typically less than ten \cite{Han2016_DABARA,han_channel_2016,Yu2020}, and this number decreases to one with high-gain antennas or massive antenna arrays.
Moreover, the gap between the LoS and NLoS path gains is significant, where the first- and second-order reflected paths are attenuated by an average of $\unit[5\!-\!10]{dB}$ and more than $\unit[15]{dB}$, respectively \cite{priebe2013stochasticchannelmodel,Han2015MR}.
Therefore, the THz channels are LoS-dominant and NLoS-assisted. THz signals are also more susceptible to self- and dynamic- blockage effects than mmWave signals \cite{shafie2020multi}. Mitigating blockage requires deploying THz BSs at sufficiently high altitudes, which further motivates the 3D channel model of this work.

We adopt the SV model, with some modifications, for indoor THz MP channels \cite{lin2015indoor}. We assume that MP components arrive in clusters \eqref{eq:ch_delay_domain}, each of which consists of several rays, where the total delay of MP components $\tau_{c,\ell}^{\NLoS}\!=\!T_c^{\NLoS}\!+\!t_{c,\ell}^{\NLoS}$, and $T_c^{\NLoS}$ and $t_{c,\ell}^{\NLoS}$ denote the cluster and ray within cluster ToAs, respectively.
The arrival times are exponentially distributed random variables conditioned on the ToA of the previous cluster/ray:
\begin{equation}
    \label{eq:cluster_toa}
    \PP(T_{c}^{\NLoS}|T_{c-1}^{\NLoS})=\Lambda e^{-\Lambda(T_c^{\NLoS}-T_{c-1}^{\NLoS})},\ T_c^{\NLoS}>T_{c-1}^{\NLoS},
\end{equation}
\begin{equation}
    \label{eq:rays_toa}
    \PP(t_{c,\ell}^{\NLoS}|t_{c,\ell-1}^{\NLoS})=\dot{\Lambda} e^{-\dot{\Lambda}(t_{c,\ell}^{\NLoS}-t_{c,\ell-1}^{\NLoS}}),\ t_{c,\ell}^{\NLoS}>t_{c,\ell-1}^{\NLoS},
\end{equation}
where $\Lambda$ and $\dot{\Lambda}$ represent the cluster and ray arrival rates, respectively. The inter/intra-cluster ToA parameters are frequency-dependent and sensitive to building materials (due to tiny wavelengths) \cite{PriebeAoADT,priebe2013stochasticchannelmodel}.
Consequently, the number of clusters and rays per cluster, $N_{\clu}$ and $N_{\ray}^{c}$, follow two independent Poisson processes of rates $\Lambda$ and $\dot{\Lambda}$, respectively. Furthermore, the average MP gain of \eqref{eq:NLoS_pg} follows a double exponential decay profile
\begin{equation}
    \label{eq:alpha_nlos_pg}
    \EE[|\alpha_{c,\ell}^{\NLoS}(f_{k},d)|^2]=|\alpha^{\LoS}(f_{k},d)|^2 e^{-\frac{T_{c}^{\NLoS}}{\Gamma}} e^{-\frac{t_{c,\ell}^{\NLoS}}{\dot{\Gamma}}},
\end{equation}
where ${\Gamma}$ and ${\dot{\Gamma}}$ denote the cluster and ray exponential decay factors (also frequency- and material-dependent). 

The total AoDs/AoAs of MPs is expressed as the summation of the ray and the cluster AoDs/AoAs. Hence, we have:
\begin{equation}
\label{eq:AoA_AoD}
\begin{array}{ll}
\phi_{c,\ell}^{(\mathrm{t})}=\Phi_{c}^{(\mathrm{t})}+\varphi_{c,\ell}^{(\mathrm{t})}, & \phi_{c,\ell}^{(\mathrm{r})}=\Phi_{c}^{(\mathrm{r})}+\varphi_{c,\ell}^{(\mathrm{r})}, \\
\theta_{c,\ell}^{(\mathrm{t})}=\Theta_{c}^{(\mathrm{t})}+\hat{\theta}_{c,\ell}^{(\mathrm{t})}, & \theta_{c,\ell}^{(\mathrm{r})}=\Theta_{c}^{(\mathrm{r})}+\hat{\theta}_{c,\ell}^{(\mathrm{r})},
\end{array}
\end{equation}
where $\Phi_{c}^{(\mathrm{t})}$/$\Phi_{c}^{(\mathrm{r})}$ and $\Theta_{c}^{(\mathrm{t})}$/$\Theta_{c}^{(\mathrm{r})}$ are the azimuth and elevation cluster AoDs/AoAs, uniformly distributed over $(-\pi, \pi]$ and $\left[-\frac{\pi}{2}, \frac{\pi}{2}\right]$, respectively, and $\varphi_{c,\ell}^{(\mathrm{t})}/\varphi_{c,\ell}^{(\mathrm{r})}$ and $\hat{\theta}_{c,\ell}^{(\mathrm{t})}/\hat{\theta}_{c,\ell}^{(\mathrm{r})}$ are the azimuth and elevation ray AoDs/AoAs that follow a zero-mean second-order Gaussian mixture distribution ($\GMM$) 
\begin{equation}
    \label{eq:gmm_dist}
    GMM(o)=\frac{w_{1}}{\sqrt{2\pi\sigma^{2}_{1}}} e^{-\frac{o^{2}}{2\sigma_{1}^{2}}}+\frac{w_{2}}{\sqrt{2\pi\sigma^{2}_{2}}} e^{-\frac{o^{2}}{2\sigma_{2}^{2}}},
\end{equation}
truncated to relevant ranges (low THz angular spread \cite{LinAoSa}); typical $\sigma^2_1$, $\sigma^2_2$, $w_1$, and $w_2$ values are reported in \cite{PriebeAoADT}.

Several models for cluster parameters exist in the literature. For instance, an extensive model for both angular and RMS delay spreads is proposed in \cite{Priebe2014}. In \cite{PriebeAoADT}, the ToA is modeled as a paraboloid function that relies on AoA information. In \cite{ju20203}, $N_{\clu}$ follows a Poisson distribution in NLoS indoor scenarios and a uniform distribution in LoS scenarios, whereas $N_{\ray}^{c}$ follows a composite of a dirac function and a discrete exponential distribution, (based on measurement data valid for the mmWave and sub-THz bands). Furthermore, the AoDs/AoAs adopt the spatial lobe concept to represent the main directions of departure/arrival, where $\Phi_{c}$/$\Phi_{c}$ follow a uniform random variable, and $\hat{\theta}_{c,\ell}/\hat{\theta}_{c,\ell}$ follow a zero-mean normal random variable. Nevertheless, in \cite{Wu2021}, the path loss due to small-scale fading follows a Gamma distribution in the presence of the LoS path and an exponential distribution in the absence of LoS path, where all model parameters are distance-dependent.

\subsection{Molecular Absorption}
\label{sec:absorption}

The molecular absorption loss is a function of the carrier frequency and the communication distance and is mainly due to water vapor molecules~\cite{Han2018UMMIMO,Han2016M-W-D-A,Han2016_DABARA}. The high attenuation absorption peaks due to excited molecule vibrations at specific THz resonant frequencies result in multiple transmission windows~\cite{zakrajsek2017design}, each having a bandwidth that shrinks with communication distance. Moreover, higher gas mixing ratios and densities result in stronger and wider spectral absorption peaks. Molecular absorption thus results in frequency-selectivity even in LoS scenarios. The molecular absorption coefficient, $\mathcal{K}(f)$, represents a unique THz fingerprint for each gas, $g$, and isotopologue, $i$. By analogy with ~\cite{Jornet2011nanothz}, $\mathcal{K}(f)$ is expressed in \eqref{eq:abs_coeff},
\begin{figure*}[ht]
\begin{equation}
\begin{aligned}
\label{eq:abs_coeff}
    & \mathcal{K}(f) = \sum_{i,g} N_A\frac{P}{P_0} \frac{T_{\mathrm{STP}}}{T} \frac{P}{RT} \xi^{(i,g)}  S^{(i,g)}(T) \left(\frac{f}{f_{c}^{(i,g)}}\right)^2 \frac{\text{tanh}\frac{\hbar c_{0}f}{2K_{B}T}}
    {\text{tanh}\frac{\hbar c_{0}f_{c}^{(i,g)}}{2K_{B}T}}\frac{1}{\pi}\left[\frac{\alpha_L^{(i,g)}}{(f-f_{c}^{(i,g)})^2+(\alpha_L^{(i,g)})^2}+\frac{\alpha_L^{(i,g)}}{(f+f_{c}^{(i,g)})^2+(\alpha_L^{(i,g)})^2}\right],
\end{aligned}
\end{equation}
\end{figure*}
where $T$ is the system temperature (in $\mathrm{Kelvin}$), $T_0$ is reference temperature ($\unit[296.0]{Kelvin}$), $T_{\mathrm{STP}}$ is a temperature at standard pressure ($\unit[273.15]{Kelvin}$), $P$ is the system pressure (in $\mathrm{atm}$), $P_0$ is the reference pressure ($\unit[1]{atm}$), $\hbar$ is the Planck constant, $K_B$ is the Boltzmann constant, $R$ is the gas constant, and $N_A$ is the Avogadro constant.
Furthermore, $\xi^{(i,g)}$, $S^{(i,g)}(T)$, $f_{c}^{(i,g)}$, and $\alpha_L^{(i,g)}$ are respectively the mixing ratio, line intensity (in $\mathrm{Hz\ m^2/molecule}$), resonant frequency (in $\mathrm{Hz}$), and Lorentz half-width (in $\mathrm{Hz}$) of isotopologue $i$ of gas $g$. In addition,
\begin{equation}
\label{eq:line_int_temp}
    S^{(i,g)}(T) = S_0^{(i,g)} \frac{\mho(T_0)}{\mho(T)} \frac{e^{-\frac{\hbar E_L^{i}}{K_B T}}}{e^{-\frac{\hbar E_L^{i}}{K_B T_0}}} \left(\frac{1-e^{-\frac{\hbar f_{c}^{(i,g)}}{K_B T}}}{1-e^{-\frac{\hbar f_{c}^{(i,g)}}{K_B T_0}}}\right),
\end{equation}
\begin{equation}
\label{eq:resonant_freq}
    f_{c}^{(i,g)} = f_{c0}^{(i,g)}+\varsigma^{(i,g)}\frac{P}{P_{0}} ,
\end{equation}
\begin{equation}
\label{eq:Lorentz_hw}
    \alpha_L^{(i,g)} = \left[\left(1-\xi^{(i,g)}\right)\alpha_0^{(\mathrm{air})}+\xi^{(i,g)}\alpha_0^{(i,g)}\right]\left(\frac{P}{P_{0}}\right)\left(\frac{T_0}{T}\right)^\iota.
\end{equation}
In~\eqref{eq:line_int_temp}, $E_L^{i}$ is the lower state energy of the transition of absorbing species, and the partition function $\mho(T)$ and its definitions are found in \cite{gordon2017hitran2016} (Appendix A). In~\eqref{eq:resonant_freq}, $f_{c0}^{(i,g)}$ is the zero-pressure position of the resonance and $\varsigma^{(i,g)}$ is the linear pressure shift.
Other parameters such as the line intensity for the reference temperature, $S_0^{(i,g)}$, the air- and self-broadened half-widths, $\alpha_0^{(\mathrm{air})}$ and $\alpha_0^{(i,g)}$, and the temperature broadening coefficient, $\iota$, are directly retrieved from the high-resolution transmission molecular absorption (HITRAN) database \cite{gordon2017hitran2016}.

Although the HITRAN-based absorption model using radiative transfer theory \cite{Jornet2011nanothz} is the most accurate, it has a high computational complexity (increased simulation time), and it is hard to track analytically. In light of this observation, two works~\cite{Kokkoniemi2018Simplified, kokkoniemi2020lineofsight} have proposed alternative approximate absorption coefficient calculations~\cite{Jornet2011nanothz}, valid at sub-THz bands. The approximations are achieved by focusing on the dominant water vapor effect only. In the first approximation \cite{Kokkoniemi2018Simplified}, valid over $\unit[275-400]{GHz}$, the absorption coefficient is expressed as the sum of two functions and an equalization factor:
\begin{align}
\label{eq:absor_coeff_approx1}
    \mathcal{K}_{\apx_1}(f)&=\Pi_{1}(f, \mu_{\htwoo})+\Pi_{2}(f, \mu_{\htwoo})+\kappa_{\apx_1}(f),\\
    \Pi_{1}(f, \mu_{\htwoo})&=\frac{V_1(\mu_{\htwoo})}{V_2(\mu_{\htwoo})+\left(\frac{f}{100 c_0}-\upsilon_{1}\right)^{2}},\\
    \Pi_{2}(f, \mu_{\htwoo})&=\frac{V_3(\mu_{\htwoo})}{V_4(\mu_{\htwoo})+\left(\frac{f}{100 c_0}-\upsilon_{2}\right)^{2}},\\
    \kappa_{\apx_1}(f)&=\bar{\rho}_{1} f^{3}+\bar{\rho}_{2} f^{2}+\bar{\rho}_{3} f+\bar{\rho}_{4},
\end{align}
where ${\mu_{\htwoo}}$ is the volume mixing ratio of water vapor. The rest of the coefficients and functions are defined in \cite{Kokkoniemi2018Simplified}. In the second approximation \cite{kokkoniemi2020lineofsight}, valid over a wider range $\unit[100-450]{GHz}$, where the expansion of the frequency range gives rise to more absorption spikes, the absorption coefficient is expressed as the sum of six elements and an equalization factor:
\begin{align}
\label{eq:absor_coeff_approx2}
    \mathcal{K}_{\apx_2}(f) &= \sum_{e=1}^{6}\hat{\Pi}_{e}(f, \mu_{\htwoo})+\kappa_{\apx_2}(f,\mu_{\htwoo}),\\
    \hat{\Pi}_{e}(f, \mu_{\htwoo})&=\frac{\hat{V}_{e}(\mu_{\htwoo})}{\bar{V}_{e}(\mu_{\htwoo})+\left(\frac{f}{100 c_0}-\hat{\upsilon}_{e}\right)^{2}},\\
    \kappa_{\apx_2}(f,\mu_{\htwoo})&=\frac{\mu_{\htwoo}}{\hat{\rho}_{1}}\left(\hat{\rho}_{2}+\hat{\rho}_{3}f^{\hat{\rho}_{4}}\right).
\end{align}
See~\cite{kokkoniemi2020lineofsight} for definitions of coefficients and functions. In both approximations, we have ${\mu_{\htwoo}}\!=\!\frac{\zeta}{100}\frac{P^{\ast}_{\omega}(T,P)}{P}$ at relative humidity (RH), $\zeta$, where $P^{\ast}_{\omega}(T,P)$ is the saturated water vapor partial pressure.
Both approximations have high accuracy for up to $\unit[1]{Kilometer}$ links under standard atmospheric conditions. We incorporate the two approximations alongside the HITRAN-based model into TeraMIMO. The HITRAN-based model is favored in joint sensing and communications settings, where exact knowledge of all medium components is sought.

\subsection{Equivalent Array Response: Beamsteering and Beamforming}
\label{sec:sa_ear_sv_bm}

We assume an AoSA antenna configuration and hybrid beamforming to reduce the complexity of the transceiver. Precoding is applied at the baseband per subcarrier. Analog beamforming is configured independently in each SA. Without loss of generality, for convenience, we adopt an equivalent array response at the Tx and Rx sides \cite{lin2015indoor}, denoted by $\mathcal{A}_{eq}^{\mathrm{(t)}}$ and $\mathcal{A}_{eq}^{\mathrm{(r)}}$, respectively. The SA array response vectors serve as beamsteering vectors that can be further expressed as a function of the transmit and receive mutual coupling matrices $\mbf{M}^{(\mathrm{t})},\mbf{M}^{(\mathrm{r})}\!\in\!\RR^{\bar{Q}\times\bar{Q}}$
\begin{align}
    \label{eq:SA_str_vec_mutcoup}
    \mbf{a}^{(\mathrm{t})}(\mbf{\Phi}^{(\mathrm{t})})=\mbf{M}^{(\mathrm{t})}\mbf{a}_{\svv}^{(\mathrm{t})}(\mbf{\Phi}^{(\mathrm{t})}),&\\
    \mbf{a}^{(\mathrm{r})}(\mbf{\Phi}^{(\mathrm{r})})=\mbf{M}^{(\mathrm{r})}\mbf{a}_{\svv}^{(\mathrm{r})}(\mbf{\Phi}^{(\mathrm{r})}).
\end{align}
By setting $\mbf{M}^{(\mathrm{t})}\!=\!\mbf{M}^{(\mathrm{r})}\!=\!\mbf{I}_{\bar{Q}}$, we neglect the effect of mutual coupling. This is not a stringent assumption, especially in plasmonic scenarios where $\delta_m^{(\mathrm{t})},\delta_n^{(\mathrm{t})}, \delta_m^{(\mathrm{r})},\delta_n^{(\mathrm{r})}\!\geq\! \lambda_{\spp}$, where $\lambda_{\spp}$ is the surface plasmon polariton (SPP) wavelength separation, which renders mutual coupling negligible ($\lambda_{\spp}$ is much smaller than the free-space wavelength, $\lambda$) \cite{Han2018UMMIMO}. We adopt in TeraMIMO ideal SA beamsteering vectors corresponding to the LoS case with perfect alignment. The steering vector is thus decided by the array structure and the AoD/AoA, and can be expressed as
\begin{equation}
    \mbf{a}_{\svv}(\mbf{\Phi}) = \frac{1}{\sqrt{\bar{Q}}}[a(1), \cdots, a(\bar{q}),\cdots,a(\bar{Q})]^\Tr.
    \label{eq:steering_vector_ae}
\end{equation}
The $\nth{\bar{q}}$ element of $\mbf{a}_{\svv}(\mbf{\Phi})$ at subcarrier frequency $f_{k}$ is
\begin{equation}
    a(\bar{q}) = e^{j\frac{2\pi}{\lambda_{k}}\Psi_{\bar{q}}(\mbf{\Phi})},
    \label{eq:steering_vector_perae_comp}
\end{equation}
where $\Psi_{\bar{q}}$ is the phase shift at the $\nth{\bar{q}}$ AE, expressed as
\begin{equation}
\begin{split}
    \Psi_{\bar{q}}(\mbf{\Phi})& = (\dot{\mbf{p}}_{q,\bar{q}})^{\Tr} \mbf{t}\\
    & = \dot{p}_{x}^{q,\bar{q}}\cos(\phi)\sin(\theta) \!+\!  \dot{p}_{y}^{q,\bar{q}}\sin(\phi)\sin(\theta) \!+\! \dot{p}_{z}^{q,\bar{q}}\cos(\theta),
    \label{eq:ae_phshift_3d}
\end{split}
\end{equation} 
and where $\dot{\mbf{p}}_{q,\bar{q}}=\mbf{p}_{q,\bar{q}} - \mbf{p}_{q}=[\dot{p}_{x}^{q,\bar{q}},\dot{p}_{y}^{q,\bar{q}},\dot{p}_{z}^{q,\bar{q}}]^{\Tr}$ are the local 3D coordinates of AEs.
In global coordinates we have
\begin{equation}
\begin{split}
    \Psi_{\bar{q}}(\tilde{\mbf{\Phi}}) &= (\tilde{{\mbf{p}}}_{q,\bar{q}}-\tilde{{\mbf{p}}}_{q})^\Tr \tilde{\mbf{t}}\\
    &= \left(\mbf{R}({\mbf{p}}_{q,\bar{q}}-{\mbf{p}}_{q})\right)^\Tr\mbf{R}\mbf{t} 
    =(\dot{\mbf{p}}_{q,\bar{q}})^{\Tr} \mbf{t}=\Psi_{\bar{q}}(\mbf{\Phi}).
    \label{eq:from_loc_to_glob_pos}
\end{split}
\end{equation}

When adopting the beamforming angles, $\mbf{\Phi}_{0}$, required to direct the array response to the target direction angles, the beamforming vector is defined as
\begin{equation}
\begin{split}
    \hat{\mbf{a}}(\mbf{\Phi}_{0}) &= [\hat{a}(1), \cdots, \hat{a}(\bar{q}),\cdots,\hat{a}(\bar{Q})]^\Tr, \\
    \hat{a}(\bar{q}) &= e^{-j\frac{2\pi}{\lambda_{k}}\Psi_{\bar{q}}(\mbf{\Phi}_{0})}.
    \label{eq:idealbm_steering_vector_ae}
\end{split}
\end{equation}
The target direction is within the antenna sector range, not necessarily the LoS path's direction, and may differ across SAs or NLoS rays. In TeraMIMO, we do not implement a beamforming solution such as a beamsteering codebook search to overcome the THz propagation constraints; we assume the target direction to be that of the LoS direction per SA (we do not normalize by $\sqrt{\bar{Q}}$ in \eqref{eq:idealbm_steering_vector_ae}). The beamforming angles are similarly expressed as
\begin{equation}
\begin{split}
    \Psi_{\bar{q}}&(\mbf{\Phi}_{0}) = (\dot{\mbf{p}}_{q,\bar{q}})^\Tr \mbf{t} \\
    & \!=\! \dot{p}_{x}^{q,\bar{q}}\cos(\phi_0)\sin(\theta_0) \!+\!  \dot{p}_{y}^{q,\bar{q}}\sin(\phi_0)\sin(\theta_0) \!+\! \dot{p}_{z}^{q,\bar{q}}\cos(\theta_0).
    \label{eq:bm_phshift_3d}
\end{split}
\end{equation}
The equivalent array response can then be expressed as
\begin{equation}
    \mathcal{A}_{eq}(\mbf{\Phi},\mbf{\Phi}_{0})= \frac{1}{\sqrt{\bar{M}\bar{N}}}\sum^{\bar{M}}_{\bar{m}=1}\sum^{\bar{N}}_{\bar{n}=1}e^{j\frac{2\pi}{\lambda_k}(\Psi_{\bar{m},\bar{n}}(\mbf{\Phi})-\Psi_{\bar{m},\bar{n}}(\mbf{\Phi}_{0}))},
    \label{eq:Aeq_3darray}
\end{equation}
where $\Psi_{\bar{m},\bar{n}}(\mbf{\Phi})$ and $\Psi_{\bar{m},\bar{n}}(\mbf{\Phi}_{0})$ are obtained from~\eqref{eq:ae_phshift_3d} and~\eqref{eq:bm_phshift_3d}, respectively.
Note that with MC schemes, the phase shifts are divided by $\lambda_{k}$ instead of $\lambda_{c}$. We assume the use of wideband THz PSs with no phase uncertainty; the beamformer phase delays are frequency-independent and do not add to the beam split effect. For the particular case of a uniform planar array (UPA) with antenna positions as defined in~\eqref{eq:local_ae_position}, the AE phase shifts are simplified as
\begin{equation}
\begin{split}
        \Psi_{\bar{q}}(\mbf{\Phi}) &= \dot{p}_{y}^{q,\bar{q}}\sin(\phi)\sin(\theta)+ \dot{p}_{z}^{q,\bar{q}}\cos(\theta),\\
        \Psi_{\bar{q}}(\mbf{\Phi}_{0}) &= \dot{p}_{y}^{q,\bar{q}}\sin(\phi_0)\sin(\theta_0)+ \dot{p}_{z}^{q,\bar{q}}\cos(\theta_0),
    \label{eq:ae_bm_phshift_upa}
\end{split}
\end{equation}
and the equivalent array response can be expressed as
\begin{equation}
   \mathcal{A}_{eq}(\mbf{\Phi},\mbf{\Phi}_{0})=  \frac{1}{\sqrt{\bar{M}\bar{N}}}\frac{\sin(\bar{M}\Omega_{\bar{M}})}{\sin(\Omega_{\bar{M}})}\frac{\sin(\bar{N}\Omega_{\bar{N}})}{\sin(\Omega_{\bar{N}})},
\label{eq:Aeq_upaarray}
\end{equation}
where
\begin{equation}
\begin{split}
    \Omega_{\bar{M}} &=  \frac{\pi\delta_m}{\lambda_k}[\cos(\theta) - \cos(\theta_0)],\\
    \Omega_{\bar{N}} &=  \frac{\pi\delta_n}{\lambda_k}[\sin(\phi)\sin(\theta) - \sin(\phi_0)\sin(\theta_0)].
    \label{eq:Omega_MN}
\end{split}
\end{equation}
As the knowledge on THz devices matures, the ideal beamforming assumption could be modified to account for more practical THz-band phase uncertainties in wideband PSs, mainly due to imperfections in materials operating at high frequencies \cite{lin2015indoor}. Such uncertainties result in beam pointing errors which could severely degrade the overall system performance due to losses in equivalent array gains and capacity, especially with small SA sizes. Denoting by $\Delta\Psi_{\bar{q}^{(\mathrm{t})}}(\mbf{\Phi},f)$ and $\Delta\Psi_{\bar{q}^{(\mathrm{r})}}(\mbf{\Phi},f)$ the frequency-dependent random phase errors at the Tx and Rx, respectively, the beamforming angles can be modified as
\begin{equation}
\begin{split}
    \tilde{\Psi}_{\bar{q}^{(\mathrm{t})}}(\mbf{\Phi}_{0},f) &= \Psi_{\bar{q}^{(\mathrm{t})}}(\mbf{\Phi}_{0}) + \Delta\Psi_{\bar{q}^{(\mathrm{t})}}(\mbf{\Phi}_{0},f),\\
    \tilde{\Psi}_{\bar{q}^{(\mathrm{r})}}(\mbf{\Phi}_{0},f) &= \Psi_{\bar{q}^{(\mathrm{r})}}(\mbf{\Phi}_{0}) + \Delta\Psi_{\bar{q}^{(\mathrm{r})}}(\mbf{\Phi}_{0},f).    
    \label{eq:ps_with_uncertaintities}
\end{split}
\end{equation}

\subsection{Antenna Gains}
\label{sec:ant_g}

Directional antennas are essential for overcoming the high THz propagation losses. Towards this end, antenna models are required to obtain the antenna gains, $G^{(\mathrm{t})}$ and $G^{(\mathrm{r})}$, at the Tx and Rx, respectively. We adopt the simplified ideal sector model (ISM)~\cite{lin2015indoor}:
\begin{equation}
G=\left\{
    \begin{array}{c l}\sqrt{G_0},& \forall {\phi}\in[\phi_{\mathrm{min}},\phi_{\mathrm{max}}],\forall{\theta}\in[\theta_{\mathrm{min}},\theta_{\mathrm{max}}],\\0,&\mathrm{otherwise}.\end{array}
\right.
 \label{eq:antenna_gain_sector_model}
\end{equation}
We apply \eqref{eq:antenna_gain_sector_model} to both LoS and NLoS components, with the corresponding azimuth and elevation angles. We assume antennas with perfect radiation efficiency, i.e., conduction, dielectric, and reflection efficiency equal to one. For highly directional antennas, $G_0$ can be approximated as~\cite{balanis2016antenna,xia2019link}
\begin{equation}
    G_0 = \frac{4\pi}{\psi_A} \approx \frac{4\pi}{\psi_{\phi}\psi_{\theta}},
    \label{eq:antenna_gain_approx}
\end{equation}
where $\psi_A$ is the beam solid angle, and $\psi_{\phi}$ and $\psi_{\theta}$ are the half-power beamwidths (HPBWs) in the azimuth and elevation planes, respectively. Here, we have $\phi_{\mathrm{min}}\!=\!-\frac{\psi_{\phi}}{2}$, $\phi_{\mathrm{max}}\!=\!\frac{\psi_{\phi}}{2}$, $\theta_{\mathrm{min}}\!=\!\frac{\pi}{2}\!-\!\frac{\psi_{\theta}}{2}$, and $\theta_{\mathrm{max}}\!=\!\frac{\pi}{2}\!+\!\frac{\psi_{\theta}}{2}$, for compatibility with our system model, where a phase shift of $\frac{\pi}{2}$ for ray AoD/AoA distribution is required to ensure that elevation angles are in the range $\![0,\pi]$.
Note that these sectors are usually small under antenna directivity. For example, HPBW azimuth/elevation-plane angles $\psi_\phi\!=\!\psi_\theta\!=\!27.7^{\circ}$ only result in a $\unit[17.3]{dBi}$ gain \cite{xia2019link}; much higher gains are required in medium-distance THz communications.

\subsection{Time-variant Channel Model}
\label{sec:tvch_domain}

When the channel coherence time, $T_{\coh}$, is smaller than the symbol time, $T_{\sym}$, the channel cannot be considered TIV. We assume the channel impulse response variations to be caused by a Doppler frequency, induced by the relative motion between Tx and Rx, where either can be moving with speed $\vartheta$ along the horizontal axis.
The equivalent per-SA TV-FS time/delay domain channel representation at time $t$ is expressed as
\begin{equation}
\begin{split}
    \label{eq:ch_tv_delay_domain}
    \hat{h}_{q^{(\mathrm{r})},q^{(\mathrm{t})}}&(t,\tau)=\alpha^{\LoS}(t)G^{(\mathrm{t})}(\mbf{\Phi}^{(\mathrm{t})})G^{(\mathrm{r})}(\mbf{\Phi}^{(\mathrm{r})})\\
    &\times\mathcal{A}_{eq}^{(\mathrm{r})}(\mbf{\Phi}^{(\mathrm{r})},\mbf{\Phi}_{0}^{(\mathrm{r})})\mathcal{A}_{eq}^{(\mathrm{t})^{\Tr}}(\mbf{\Phi}^{(\mathrm{t})},\mbf{\Phi}_{0}^{(\mathrm{t})})\mathrm{dirac}(t-\tau^{\LoS}(t))\\
    &+\sum_{c=1}^{N_{\clu}}\sum_{\ell=1}^{N_{\ray}^{c}}\alpha_{c,\ell}^{\NLoS}(t)G^{(\mathrm{t})}(\mbf{\Phi}_{c,\ell}^{(\mathrm{t})})G^{(\mathrm{r})}(\mbf{\Phi}_{c,\ell}^{(\mathrm{r})})\\
    &\times\mathcal{A}_{eq}^{(\mathrm{r})}(\mbf{\Phi}_{c,\ell}^{(\mathrm{r})},\mbf{\Phi}_{0}^{(\mathrm{r})})\mathcal{A}_{eq}^{(\mathrm{t})^{\Tr}}(\mbf{\Phi}_{c,\ell}^{(\mathrm{t})},\mbf{\Phi}_{0}^{(\mathrm{t})})\mathrm{dirac}(t-\tau^{\NLoS}_{c,\ell}).
\end{split}
\end{equation}
For the LoS component, we consider $d(t)$ to be a TV distance between Tx and Rx due to the relative motion \cite{nie2017three}. In~\eqref{eq:ch_tv_delay_domain}, we adopt Bello’s wide-sense stationary uncorrelated scattering (WSSUS) assumptions. MPs are grouped into one or more resolvable paths, in addition to unresolvable paths. The number of resolvable paths, their delays, and their second-order statistical properties are assumed to be invariant in time, and the resolvable paths are assumed statistically uncorrelated.

The auto-correlation function (ACF), for NLoS components only, and assuming unity antenna and beamsteering/beamforming gains, is defined as
\begin{equation}
    \label{eq:auto_corr_tv_delay}
    \Xi_{\hat{h}\hat{h}}(\Delta t,\tau) = \EE[\hat{h}_{q^{(\mathrm{r})},q^{(\mathrm{t})}}(t,\tau)\hat{h}^{\ast}_{q^{(\mathrm{r})},q^{(\mathrm{t})}}(t+\Delta t,\tau)],
\end{equation}
where $\Delta t$ is a time separation. We define the power spectral density (PSD), $\mathcal{S}(\nu)$, of the Doppler spectrum as the Fourier transform of the ACF of the fading gain. The Doppler spread depends on the AoA’s distribution.
One model we adopt is Clarke’s AoA model, which assumes a uniform AoA distribution over $[-\pi,\ \pi]$ in a horizontal plane. Here, the ACF is defined as \cite{iskander2008matlab}
\begin{equation}
    \label{eq:auto_corr_bessel}
    \Xi_{\Jakes}(\Delta t) = J_0(2\pi f_{\maxx}\Delta t),
\end{equation}
where $J_0$ is the Bessel function of the first kind of order zero, and the Doppler spectrum is expressed following the Jakes model as
\begin{equation}
    \label{eq:doppler_spec_jakes}
    \mathcal{S}_{\Jakes}(\nu) =\left\{
    \begin{array}{c l} \frac{1}{\pi f_{\maxx} \sqrt{1- \left(\nu/f_{\maxx}\right)}},& |\nu| \leq f_{\maxx},\\0,&\mathrm{otherwise}.\end{array}
\right.
\end{equation}

We also implement another Doppler model, the Flat Doppler spectrum, which assumes a 3D isotropic scattering environment, where the AoAs are uniformly distributed in the elevation and azimuth planes (using the GUI, users can select the model of preference; see Fig.~\ref{fig:teramimo_gui} further ahead).
In this case, the baseband normalized flat Doppler spectrum is given by
\cite{iskander2008matlab}
\begin{equation}
    \label{eq:doppler_spec_flat}
    \mathcal{S}_{\Flat}(\nu) =\left\{
    \begin{array}{c l} \frac{1}{2 f_{\maxx}},& |\nu| \leq f_{\maxx},\\0,&\mathrm{otherwise},\end{array}
\right.
\end{equation}
and its AFC is
\begin{equation}
    \label{eq:auto_corr_sinc}
    \Xi_{\Flat}(\Delta t) = \mathrm{sinc}(2 f_{\maxx}\Delta t).
\end{equation}
Note that the WSSUS assumptions might not reflect exact THz characteristics for all scenarios. The channel characteristics for different propagation delays in wideband channels differ from those of narrowband channels. For example, the AoDs/AoAs and the total number of MP components should be treated as TV parameters~\cite{yuan20153d,jiang20193}. WSSUS assumptions are valid only for very short time intervals (order of millisecond \cite{paier2008non}).
In particular, such assumptions are not suitable for a V2X THz channel model.
We adopt WSSUS assumptions in this version of TeraMIMO and consequently seek a validation process over a small time span.
Future extensions to the simulator can incorporate novel non-stationary TV channel characteristics~\cite{chen2019time,wang2020novel}.

\subsection{THz-specific Realization}
\label{sec:wideband_thz_real}
\subsubsection{Misalignment}
\label{sec:misalign_effect}

Misalignment occurs when the Tx and the Rx do not perfectly point to each other, which is highly probable with narrow THz beams~\cite{cacciapuoti2018beyond}. A THz misalignment fading model is proposed in~\cite{boulogeorgos2019analytical} based on the optical receiver's intensity fluctuation derived in~\cite{farid2007outage}. 
We consider a scenario in which only the Tx antenna experiences 2D shaking, perhaps in a fronthaul scenario where an access point is deployed in road-sides or high buildings \cite{kokkoniemi2020impact}.
We assume an Rx antenna with a circular effective area, $W_{\mathrm{r}}$, of radius $\bar{\gamma}$. We denote by $\rho$ the Tx circular beam footprint at a distance $d$, where $0 \!\leq\! \rho \!\leq\! w_t(d)$, and $w_t(d)$ is the maximum beam radius at distance $d$~\cite{boulogeorgos2019analytical}.
The PDF of the misalignment gain $b_p$ can be expressed as
\begin{equation}
    \PP_{b_p}(o) =\frac{\varepsilon^2}{W_{\mathrm{r}0}^{\varepsilon^2}}o^{\varepsilon^2-1}, \ \ \ 0\le o\le W_{\mathrm{r}0},
    \label{eq:misalignment_gain}
\end{equation}
where $W_{\mathrm{r}0}$ is the fraction of the collected power at $\varrho\!=\!0$ and $\varrho$ is the pointing error (radial distance between the Tx and Rx beams). Assuming identical and independent Gaussian distributions for both horizontal and elevation displacements \cite{farid2007outage,boulogeorgos2019analytical}, the Rx radial displacement follows a Rayleigh distribution (2D up-down and left-right independent shaking).
The equivalent beamwidth is defined as~\cite{farid2007outage}
\begin{equation}
    w_{{eq}}^2 = w_t^2\frac{\sqrt{\pi}\mathrm{erf}(v)}{2v\exp(-v^2)},
    \label{eq:misalignment}
\end{equation}
where $v\!=\!\frac{\sqrt{\pi}}{\sqrt{2}}\frac{w_r}{w_t(d)}$, $w_r$ is the radius of the receiving area, and $\mathrm{erf}(\cdot)$ stands for the error function. We then compute $\varepsilon \!=\! \frac{w_{eq}}{2\sigma_{ma}}$
and $W_{\mathrm{r}0} \!=\! \mathrm{erf}(v)^2$, where $\sigma_{ma}$ is the jitter variance. In each realization, a misalignment gain $b_p$ is generated using~\eqref{eq:misalignment_gain} and the corresponding misalignment coefficient is obtained after normalization as
\begin{equation}
    \bar{b}_p = \frac{b_p}{W_{\mathrm{r}0}}.
\end{equation}

Since \eqref{eq:misalignment_gain} is an optical-based approximate misalignment model, we complement it by another approximation. The effective radius of the receiving area, given our AoSA system model dimensions, is approximated as
\begin{equation}
    w_r = \sqrt{\frac{M^{(\mathrm{r})}\Delta_{m}^{(\mathrm{r})} N^{(\mathrm{r})}\Delta_{n}^{{(\mathrm{r})}}}{\pi}},
\end{equation}
and the Tx beam radius can be approximated as
\begin{equation}
    w_t(d) = \sqrt{\frac{\psi_Ad^2}{\pi}},
\end{equation}
with $\psi_A$ being the beam solid angle defined as~\cite{balanis2016antenna}
\begin{equation}
    \psi_A \!=\! \frac{\theta_{M^{(\mathrm{t})}}\theta_{N^{(\mathrm{t})}}\sec(\theta_0^{(\mathrm{t})})}{\!\sqrt{\!\left(\!\sin^2\!(\phi_0^{(\mathrm{t})})\!+\!\frac{\theta_{M^{(\mathrm{t})}}^2}{\theta_{N^{(\mathrm{t})}}^2}\!\cos^2\!(\phi_0^{(\mathrm{t})})\!\right)\!\left(\!\sin^2\!(\phi_0^{(\mathrm{t})}) \!+\! \frac{\theta_{N^{(\mathrm{t})}}^2}{\theta_{M^{(\mathrm{t})}}^2}\!\cos^2\!(\phi_0^{(\mathrm{t})})\!\right)\!}},
\end{equation}
where $\theta_{M^{(\mathrm{t})}}\!=\!2\arcsin\left(\frac{2.782\lambda_c}{2\pi M^{(\mathrm{t})}d}\right)$ ~\cite{balanis2016antenna} is the HPBW of an $M$-antenna uniform linear array (ULA) and $\phi_0^{(\mathrm{t})}/\theta_0^{(\mathrm{t})}$ are transmitter beamforming angles.

\subsubsection{Spherical Wave Model}
\label{sec:swm_effect}

Another essential feature of the THz channel is the SWM. The PWM applies to far-field scenarios where the distance between the Rx and the Tx is greater than or equal to the Rayleigh distance of the antenna array (the Fraunhofer region) ~\cite{bohagen2009spherical}
\begin{equation}
    d \ge \frac{2D^2}{\lambda},
\end{equation}
with $D$ being the maximum overall antenna dimension.
At lower microwave and mmWave frequencies, this distance is less than $\unit[0.5]{m}$ and $\unit[5]{m}$ for an array size of $\unit[0.1]{m}$ and an operating frequency of $\unit[6]{GHz}$ and $\unit[60]{GHz}$, respectively. However, this distance grows to approximately $\unit[40]{m}$ at $\unit[0.6]{THz}$, which is greater than most achievable THz communication distances, and hence the importance of the SWM. The SWM should be considered when the distance is within the Fresnel region \cite{balanis2016antenna}
\begin{equation}
     0.62\sqrt{\frac{D^3}{\lambda}} \le d < \frac{2D^2}{\lambda}.
\end{equation}

The SWM and PWM can be considered at both the AE-level and the SA-level. In TeraMIMO, we adopt PWM at the AE level inside a SA due to the relatively small footprints. At the SA level, we consider both PWM and SWM. PWM at the level of AEs results in a compact equivalent array response form and reduces the computational complexity. Nevertheless, the SWM can capture the curvature information when the Rx is close to the Tx, which is helpful in near-field scenarios.

The signal wave considerations of the adopted SWM are at the SA-distance and the SA-angle levels. At the SA-distance level, the distance between two SAs in the far-field can be calculated in PWM as
\begin{equation}
    d^{(\mathrm{far})}_{q^{(\mathrm{r})},q^{(\mathrm{t})}}=\norm{\tilde{\mbf{p}}_{\cent}^{(\mathrm{r})}-\tilde{\mbf{p}}_{\cent}^{(\mathrm{t})}}-\left(\mbf{p}_{{q^{(\mathrm{r})}}}^{\Tr}\mbf{t}^{(\mathrm{t},\mathrm{r})}+\mbf{p}_{{q^{(\mathrm{t})}}}^{\Tr}\mbf{t}^{(\mathrm{r},\mathrm{t})}\right).
\label{eq:sa_distance_level_PWM}
\end{equation}

Note that \eqref{eq:sa_distance_level_PWM} effectively represents SA-level beamsteering.
The distance between two SAs in the near-field SWM is accurately computed as 
\begin{equation}
        d^{(\mathrm{near})}_{q^{(\mathrm{r})},q^{(\mathrm{t})}} =\norm{\tilde{\mbf{p}}^{(\mathrm{r})}_{q^{(\mathrm{r})}}-\tilde{\mbf{p}}^{(\mathrm{t})}_{q^{(\mathrm{t})}}}.
        \label{eq:sa_distance_level_SWM}
\end{equation}

At the SA-angle level, the PWM AoD/AoA are fixed for different SAs within the AoSA following~\eqref{eq:eq_local_angle}. For an SWM, however, the angles differ between SAs as
\begin{equation}
        \mbf{\Phi}_{q^{(\mathrm{r})}, q^{(\mathrm{t})}} = 
        \begin{bmatrix}
            \phi_{q^{(\mathrm{r})}, q^{(\mathrm{t})}}\\
            \theta_{q^{(\mathrm{r})}, q^{(\mathrm{t})}}
        \end{bmatrix}=
        \begin{bmatrix}
            \atantwo\left(t_{2}^{q^{(\mathrm{r})}, q^{(\mathrm{t})}},t_{1}^{q^{(\mathrm{r})}, q^{(\mathrm{t})}}\right)\\
            \arccos\left(t_{3}^{q^{(\mathrm{r})}, q^{(\mathrm{t})}}\right)
        \end{bmatrix},
        \label{eq:sa_angle_level_SWM}
    \end{equation}
    where 
    \begin{equation}
        \mbf{t}_{q^{(r)}, q^{(t)}}
        = 
    \begin{bmatrix}
    t_{1}^{q^{(\mathrm{r})}, q^{(\mathrm{t})}}\\
    t_{2}^{q^{(\mathrm{r})}, q^{(\mathrm{t})}}\\
    t_{3}^{q^{(\mathrm{r})}, q^{(\mathrm{t})}}
    \end{bmatrix}
    =\frac{{\mbf{p}}^{(\mathrm{r})}_{q^{(r)}}-{\mbf{p}}^{(\mathrm{t})}_{q^{(t)}}}
        {\norm{{\mbf{p}}^{(\mathrm{r})}_{q^{(r)}}-{\mbf{p}}^{(\mathrm{t})}_{q^{(t)}}}}.
    \label{eq:eq_local_direction_SWM}
\end{equation}
    
Let $\mbf{H}_\mathrm{sp}$ denote the channel realized using SWM at the SA level and PWM at the AE level. Equations ~\eqref{eq:sa_distance_level_SWM} and \eqref{eq:sa_angle_level_SWM} are used for distance and angle calculations of Tx/Rx SA pairs, respectively. The AEs in an SA share the same local angle and the distances between Tx/Rx AEs are calculated using \eqref{eq:sa_distance_level_PWM}. Moreover, the channel models using PWM and SWM at both AE-/SA-levels are denoted as $\mbf{H}_\mathrm{pp}$ and $\mbf{H}_\mathrm{ss}$, respectively.
In $\mbf{H}_\mathrm{pp}$, all SAs share the same angle calculated from \eqref{eq:eq_local_angle} and the distances between Tx/Rx SAs are calculated using \eqref{eq:sa_distance_level_PWM}. However, in $\mbf{H}_\mathrm{ss}$, distances and angles are calculated using \eqref{eq:sa_distance_level_SWM} and \eqref{eq:sa_angle_level_SWM} for all AEs. By comparing the approximated channel matrix $\tilde{\mbf{H}}$ (e.g., $\mbf{H}_\mathrm{sp}$, $\mbf{H}_\mathrm{pp}$) with the benchmark matrix $\mbf{H}_\mathrm{ss}$, the channel error is computed as
\begin{equation}
    \mathrm{err}(\tilde{\mbf{H}},\mbf{H}_\mathrm{ss}) = \frac{\norm{\tilde{\mbf{H}}-\mbf{H}_\mathrm{ss}}_F}{\norm{\mbf{H}_\mathrm{ss}}_F}.
\end{equation}
\subsubsection{Beam Split}
\label{sec:beamsplit_effect}

The ultra-broadband nature of THz communications leads to large fractional bandwidths, defined as the ratio between the bandwidth and the central frequency $B_{\mathrm{fr}}\!=\!\frac{B}{f_c}$. Large fractional bandwidths result in severe performance degradation because of the beam split effect, where the THz path components squint into different spatial directions at different subcarriers, causing significant array gain loss~\cite{dai2021delay}. Such squint can be caused by the frequency-independent delays of PSs in analog beamforming when the goal is to focus the energy of a signal at a single central operating frequency $f_c$. 
PSs consist of frequency-independent components, which tune the same phase shift for signals with different frequencies, introducing a phase error in signals.
The large THz UM-MIMO arrays also introduce a beam split effect, where signal propagation delays between SAs of the same AoSA become comparable to symbol times, introducing a frequency-dependent phase shift \cite{dai2021delay}. UM-MIMO systems also generate extremely narrow beamwidths that worsen the split effect.

The resultant spatial path direction (physical propagation direction) at a specific subcarrier $f_k$ is defined as~\cite{dai2021delay}
\begin{equation}
    \label{eq:spat_dir_subc}
    \hat{\Psi}_{\bar{q}}(\mbf{\Phi},f_k) = 2 \frac{f_k}{c_0} \Psi _{\bar{q}}(\mbf{\Phi}),
\end{equation}
and when the beamforming spatial direction is computed at the center frequency, we have
\begin{equation}
    \label{eq:spat_dir_fc}
    \hat{\Psi}_{\bar{q}}(\mbf{\Phi},f_c) = 2 \frac{f_c}{c_0} \Psi _{\bar{q}}(\mbf{\Phi}).
\end{equation}
Therefore, the beam split in the angle domain is expressed as
\begin{equation}
    \label{eq:spat_dir_subc_fc}
    \hat{\Psi}_{\bar{q}}(\mbf{\Phi},f_k) = \frac{f_k}{f_c} \hat{\Psi}_{\bar{q}}(\mbf{\Phi},f_c).
\end{equation}
At THz frequencies, the difference between $f_k$ and $f_c$ cannot be ignored. The path components tend to split into very different spatial directions at different frequencies within an ultra-broadband carrier bandwidth. This beam split effect can be modeled by multiplying $\Psi_{\bar q}(\mbf{\Phi}_{0})$ in~\eqref{eq:bm_phshift_3d} with $\lambda_k/\lambda_c$ for beamforming. Then, the corresponding equivalent array response can be expressed as
\begin{equation}
    \hat{\mathcal{A}}_{eq}(\mbf{\Phi},\mbf{\Phi}_{0}) = \frac{1}{\sqrt{\bar{M}\bar{N}}}\sum^{\bar{M}}_{\bar{m}=1}\sum^{\bar{N}}_{\bar{n}=1}e^{j 2\pi\left(\frac{\Psi_{\bar{m},\bar{n}}(\mbf{\Phi})}{\lambda_k}-\frac{\Psi_{\bar{m},\bar{n}}(\mbf{\Phi}_{0})}{\lambda_c}\right)}.
    \label{eq:Aeq_3darray_beam_split}
\end{equation}
Similarly, for a UPA, the beam split effect is expressed as
\begin{equation}
\begin{split}
    \hat{\Omega}_{M} &= \frac{\pi\delta_m}{\lambda_k}\cos(\theta) - \frac{\pi\delta_m}{\lambda_c}\cos(\theta_0) ,\\
    \hat{\Omega}_{N} &= \frac{\pi\delta_n}{\lambda_k}\cos(\phi)\sin(\theta) - \frac{\pi\delta_n}{\lambda_c}\cos(\phi_0)\sin(\theta_0) .
    \label{eq:Omega_MN_split}
\end{split}
\end{equation}
Beam split can be mitigated in the digital domain of hybrid beamforming architectures \cite{han2021hybrid}.

\section{Simulator Guide}
\label{sec:guide_gui}

\begin{figure*}[ht!]
  \centering
  \captionsetup{justification=centering}
  \includegraphics[width=0.55\textwidth]{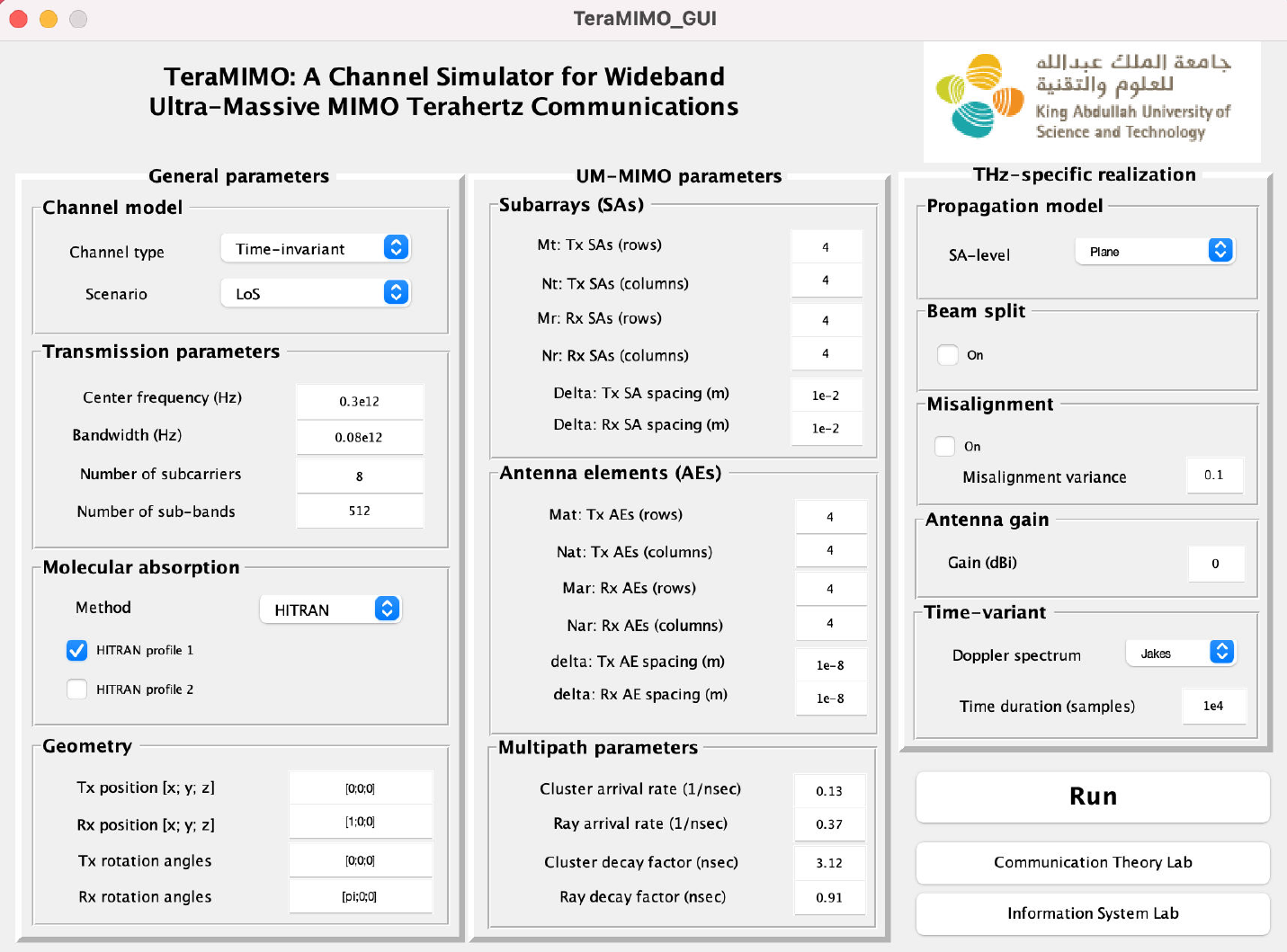}
  \caption{GUI of the TeraMIMO THz channel simulator.}
  \label{fig:teramimo_gui}
\end{figure*}

In this section, we illustrate the structure of our simulator. Note that detailed execution steps are available on GitHub~\footnote{\url{https://github.com/hasarieddeen/TeraMIMO}}. TeraMIMO supports two modes: GUI (Fig.~\ref{fig:teramimo_gui}) and MATLAB scripts.
After selecting the parameters in the GUI and running the simulation, the results are saved to the MATLAB workspace in three main structs, featuring channel information (statistics) such as $B_\coh, \tau_\rms, T_\coh,$ and $f_\maxx$ for TV channels, and $B_\coh$ and $\tau_\rms$ for the TIV channels. The exact UM-MIMO delay-domain and frequency-domain channel responses (Eqs.~\eqref{eq:ch_delay_domain} to \eqref{eq:ch_sa_freq_domain}) are also saved as outputs, alongside the molecular absorption coefficient $\mathcal{K}(f)$ (Eq.~\eqref{eq:abs_coeff}).

The input parameters can be grouped into three main categories: General parameters, UM-MIMO parameters, and THz-specific realizations, as shown in Fig.~\ref{fig:teramimo_gui}. The general parameters contain four sub-categories as follows:
\begin{enumerate}
    \item Channel model: TV or TIV, and LoS (using only the first term in Eq.~\eqref{eq:ch_sa_freq_domain}), NLoS (MP components, using only the second term in Eq.~\eqref{eq:ch_sa_freq_domain}), or LoS-dominant and NLoS-assisted (using both terms in Eq.~\eqref{eq:ch_sa_freq_domain}). Indoor scenarios are typically NLoS or LoS-dominant and NLoS-assisted, whereas outdoor scenarios are usually LoS or LoS-dominant and NLoS-assisted.
    \item Transmission parameters: Center frequency ($f_c$), bandwidth ($B$), and number of subcarriers ($K$) and sub-bands ($N_\subb$).
    \item Molecular absorption parameters: Exact (Eq.~\eqref{eq:abs_coeff}) or approximate (Eqs.~\eqref{eq:absor_coeff_approx1} or Eq.~\eqref{eq:absor_coeff_approx2}) computations. We load default sample molecule profiles to expedite exact HITRAN-based calculations. However, users can experiment with more molecule combinations by loading more “CSV” files from an offline directory.
    \item Geometry parameters: including the AoSAs centers, $\tilde{\mbf{p}}_{\cent}^{(\mathrm{t})}$ and $\tilde{\mbf{p}}_{\cent}^{(\mathrm{r})}$, rotation angles, $[\dot{\alpha}^{(\mathrm{t})},\dot{\beta}^{(\mathrm{t})},\dot{\gamma}^{(\mathrm{t})}]$ and $[\dot{\alpha}^{(\mathrm{r})},\dot{\beta}^{(\mathrm{r})},\dot{\gamma}^{(\mathrm{r})}]$ (Eq.~\eqref{eq:eq_rot_mat}), and separation distance $d_{\cent^{(\mathrm{r})},\cent^{(\mathrm{t})}}=\norm{\tilde{\mbf{p}}_{\cent}^{(\mathrm{r})}-\tilde{\mbf{p}}_{\cent}^{(\mathrm{t})}}$.
\end{enumerate}

Under UM-MIMO parameters, the user can define the number of SAs ($Q^{(\mathrm{t})}$), the number of array rows ($M^{(\mathrm{t})}$) and columns ($N^{(\mathrm{t})}$), and the distances between the row/column centers of two adjacent SAs ($\Delta_m$, and $\Delta_n$) and AEs ($\delta_m$, and $\delta_n$). The user can define MP parameters such as the cluster and ray arrival rates ($\Lambda$ and $\dot{\Lambda}$ of Eqs.~\eqref{eq:cluster_toa} and~\eqref{eq:rays_toa}) and the cluster and ray decay factors ($\Gamma$ and $\dot{\Gamma}$ of Eq.~\eqref{eq:alpha_nlos_pg}).
Moreover, under THz-specific realizations, the user decides on the SA-level propagation model (PWM or SWM), the beam split effect, and the misalignment parameters. Furthermore, the user can select antenna gains, Doppler spectra (Jakes or Flat), and simulation durations (in samples for a TV channel).

When executing MATLAB scripts rather than the GUI, the simulation parameters are configured in a single function (generate\_channel\_param\_TIV.m).
The molecular absorption coefficient is computed in a separate function (compute\_Abs\_Coef.m) that can be independently accelerated, whereas the main channel computations, for both frequency and delay domains, are executed in the main function (channel\_TIV.m). Finally, users can visualize the channel using the available plot functions (Plot\_TIV\_THz\_Channel.m).
We also provide as code output two forms of the TV channel in the time-delay (Eq.~\eqref{eq:ch_tv_delay_domain}) and the time-frequency domains.

\section{Sample Results}
\label{sec:simulations}

\begin{figure}
  \centering
  \includegraphics[width=0.48\textwidth]{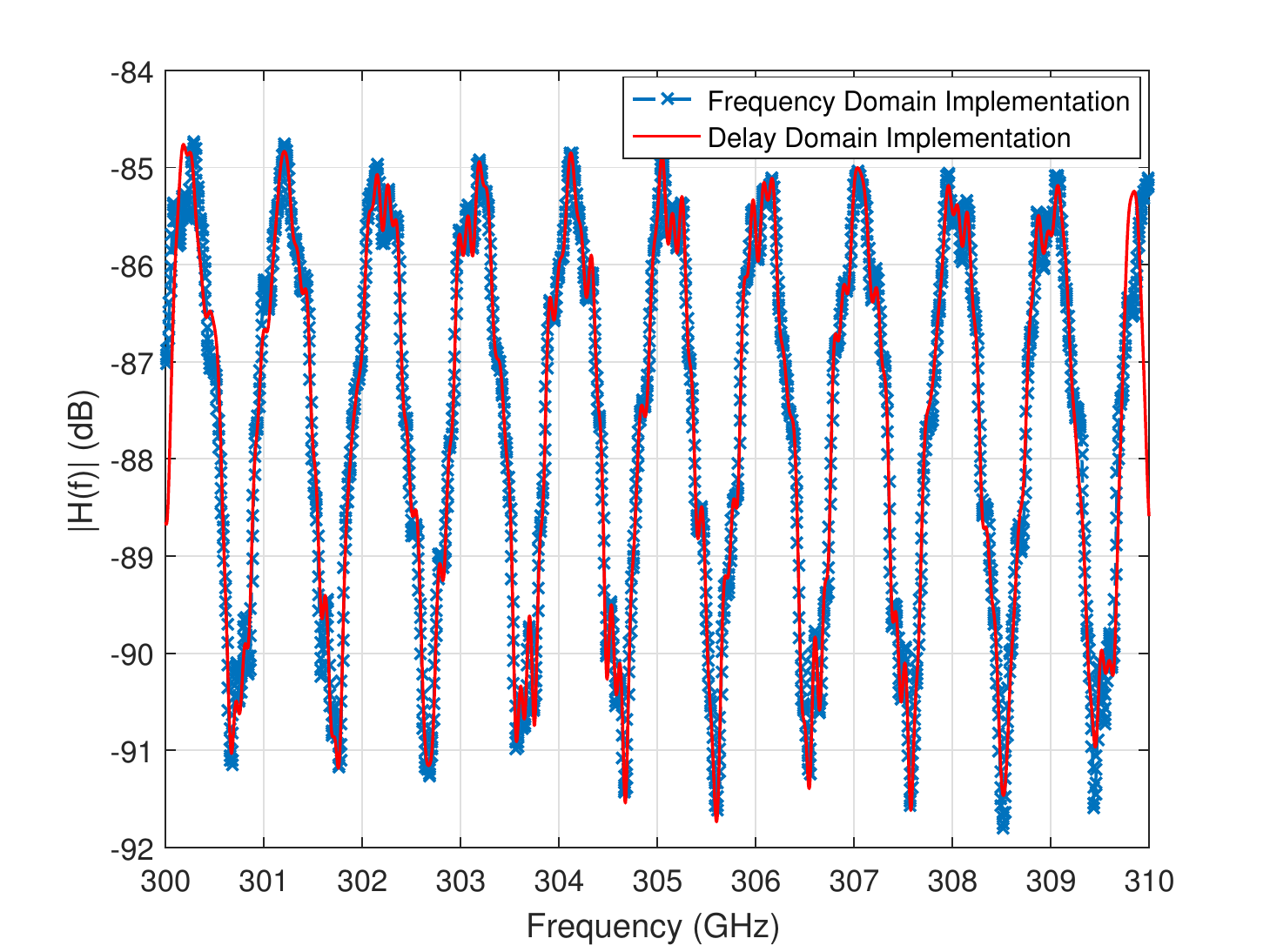}
 \caption{TeraMIMO-simulated frequency channel responses using frequency- and delay-domain implementations for $f_c\!=\!\unit[305]{GHz}$ and $B\!=\!\unit[10]{GHz}$.}
  \label{fig:compare_simulation_measurements}
\end{figure}

\begin{figure}
  \centering
  \includegraphics[width=0.48\textwidth]{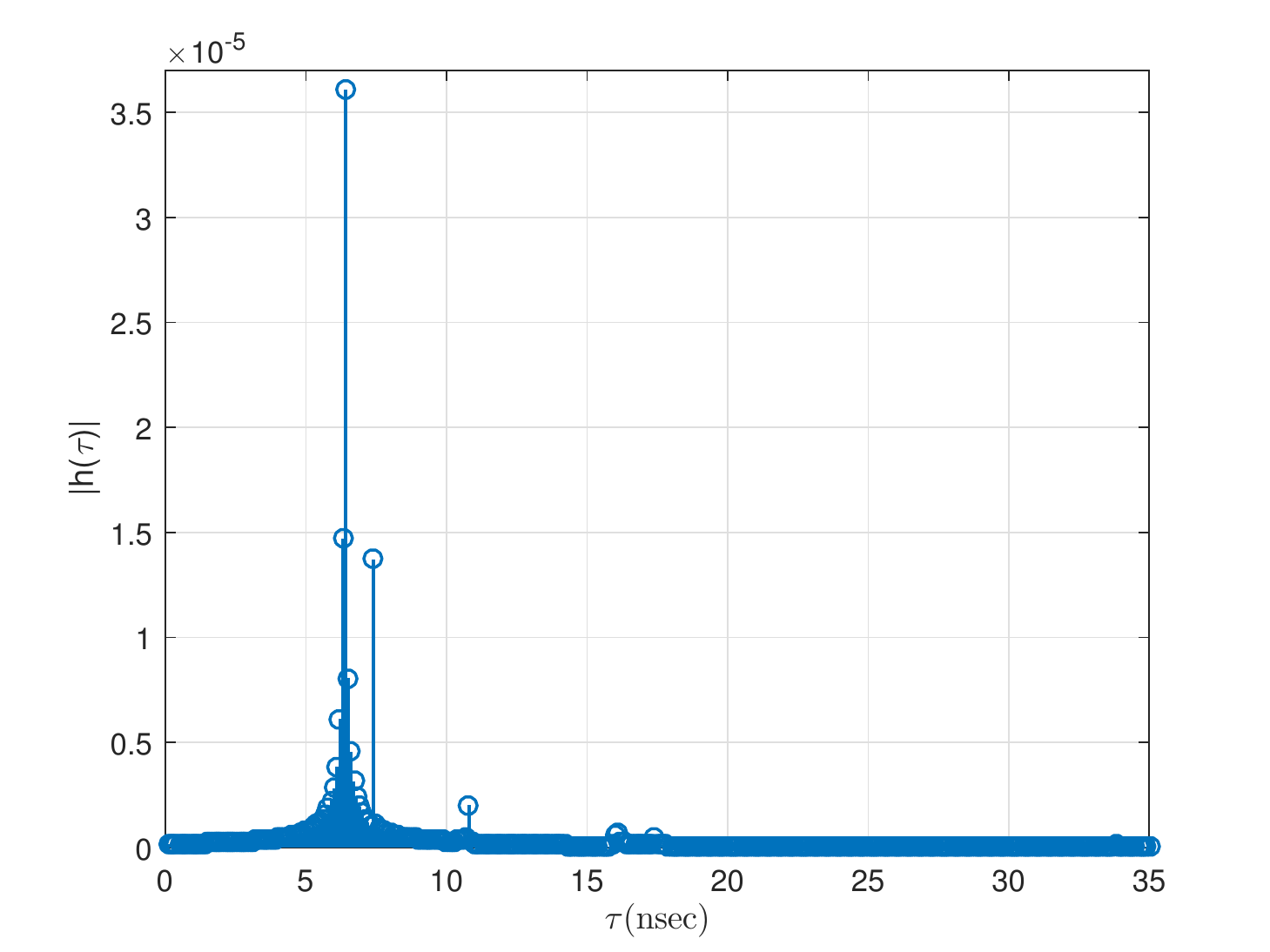}
 \caption{TeraMIMO-simulated delay-domain channel response for $f_c\!=\!\unit[305]{GHz}$ and $B\!=\!\unit[10]{GHz}$.}
  \label{fig:compare_sim_measurements_delay}
\end{figure}

TeraMIMO channel realizations in frequency and delay domains are illustrated in Fig. \ref{fig:compare_simulation_measurements} and Fig. \ref{fig:compare_sim_measurements_delay}, respectively. In Fig. \ref{fig:compare_simulation_measurements}, we also attempt to transform the delay-domain channel into the frequency domain using Fourier transform.
The results closely match those obtained from measurements (Fig. 2 in~\cite{Priebe2010_300ghz_10bw}). However, the statistical behavior of our model and the unknown material composition of room walls, gases in the medium, and exact locations of Tx and Rx are factors that cause a slight differences in the amplitude.

\begin{figure}
  \centering
  \includegraphics[width=0.48\textwidth]{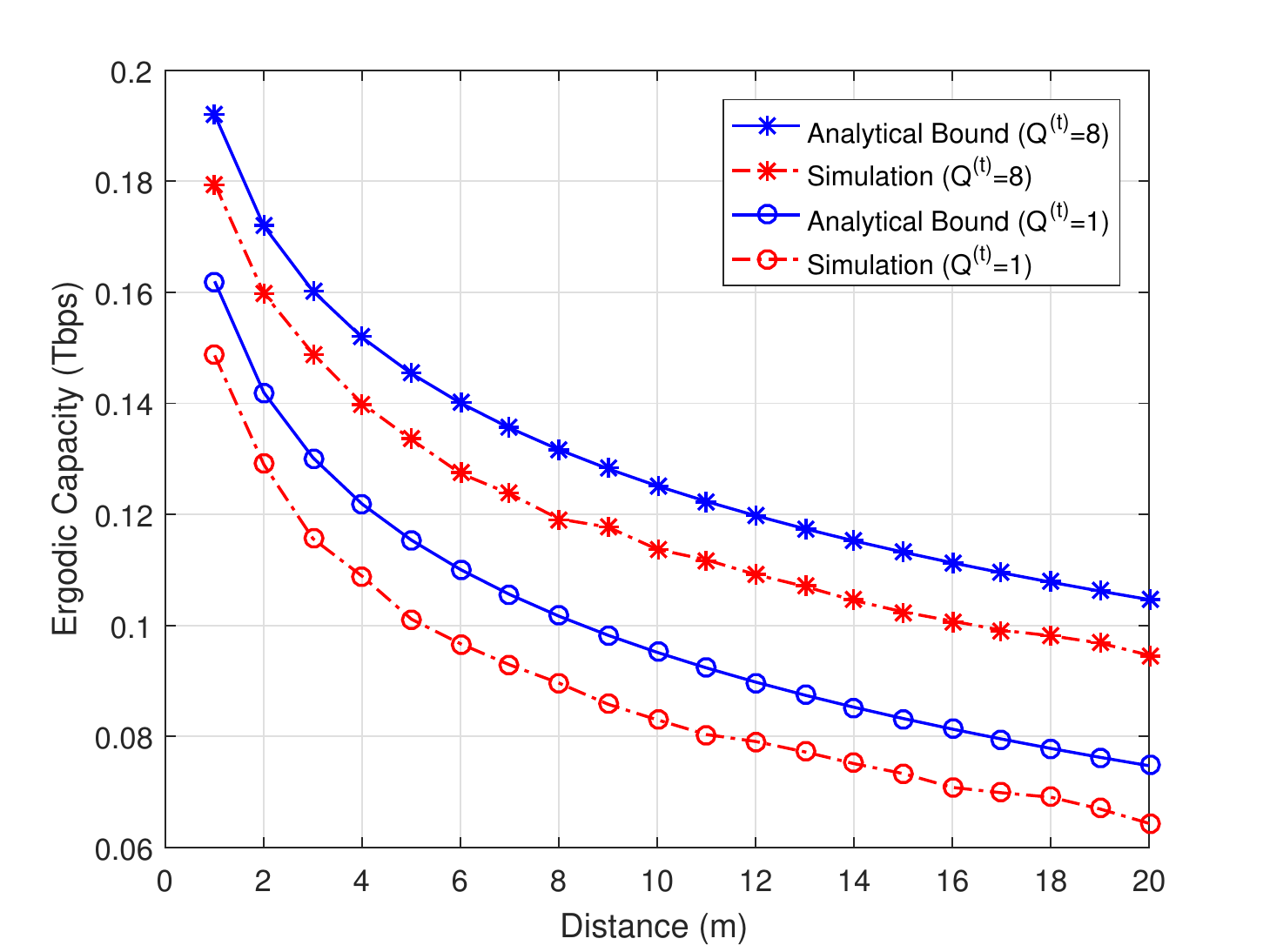}
 \caption{Simulation and analytical upper bound for the ergodic capacity in MP channels at $f_c\!=\!\unit[300]{GHz}$.}
  \label{fig:compare_ergodic_capacity}
\end{figure}

Because TeraMIMO is a stochastic simulator, it is also suitable to verify it using statistical ergodic capacity analysis, with reference analytical upper bounds. We assume (for comparison purposes) an indoor THz scenario (MP components only) and an effective multiple-input single-output (MISO) channel $\hat{\mbf{h}}$ ($Q^{(\mathrm{t})}$ SAs at the Tx and one SA at the Rx). The ergodic capacity can be expressed as~\cite{lin2015indoor} 
\begin{equation}
    \label{eq:ergodic_capacity}
    C(f,d) = \EE\left[B\log_{2}\left(1+\frac{P_{\mathrm{Tx}}}{N_0}\norm{\hat{\mbf{h}}}^2\right)\right],
\end{equation}
where $N_0$ is the noise power and $P_{\mathrm{Tx}}$ is the transmit power.
The analytical upper bound for~\eqref{eq:ergodic_capacity} is given by~\cite{lin2015indoor} 
\begin{equation}
    \label{eq:upperboudn_ergodic_capacity}
    C(f,d) \leq B\log_{2}\left(1+\frac{Q^{(\mathrm{t})}P_{\mathrm{Tx}}\bar{Q}^{(\mathrm{t})}\bar{Q}^{(\mathrm{r})}}{N_0}\beta(f,d)\right),
\end{equation}
where $\beta(f,d)$ is given by Eq. (29) in~\cite{lin2015indoor}. The simulation parameters are listed in Tab.~\ref{table:simulationpara_ergodiccap}.
All these parameters are taken from measurements-based references~\cite{PriebeAoADT,priebe2013stochasticchannelmodel,Jornet2011nanothz}. Furthermore, the GMM parameters for AoD/AoA ray distributions (Eq.~\eqref{eq:gmm_dist}) can be found in Table III in~\cite{PriebeAoADT}.
The resultant THz ergodic capacity is obtained by averaging over $5000$ channel realizations, and the number of AEs per SA is $\bar{Q}^{(\mathrm{t})}=\bar{Q}^{(\mathrm{r})}=8\times8$. The ergodic capacities for different numbers of Tx SAs are illustrated in Fig. \ref{fig:compare_ergodic_capacity} for an indoor scenario; the results tightly match those reported in Fig. 2 in \cite{lin2015indoor}.
Fig.~\ref{fig:compare_ergodic_capacity} illustrates that the ergodic capacity decreases with increasing communication distances and increases with a larger number of SAs. Besides, the analytical capacity upper bound matches the simulation results for different distances and SA sizes.
\begin{table}
\footnotesize
\centering
\caption{Ergodic capacity simulation parameters}
\begin{tabular} { l l}
 \hline
 Parameters & Values \\ [0.5ex] 
 \hline
 Operating frequency $f_c$ & $\unit[0.3]{THz}$\\
 System bandwidth $B$ & $\unit[10]{GHz}$\\ 
 Time margin & $\unit[50]{nsec}$\\
 Absorption Coefficient $\mathcal{K}$ & $\unit[0.0033]{m^{-1}}$\\ 
 Cluster arrival rate $\Lambda$ & $\unit[0.13]{nsec^{-1}}$\\
 Ray arrival rate $\dot{\Lambda}$ & $\unit[0.37]{nsec^{-1}}$\\ 
 Cluster decay factor $\Gamma$ & $\unit[3.12]{nsec}$\\
 Ray decay factor $\dot{\Gamma}$ & $\unit[0.91]{nsec}$\\ 
 Antenna Gain $G^{(\mathrm{t})}=G^{(\mathrm{r})}$ & $\unit[20]{dBi}$\\
 Tx power $P_{\mathrm{Tx}}$ & $\unit[3]{dBm}$\\
 Noise power $N_0$ & $\unit[-75]{dBm}$\\ 
  \hline
\end{tabular}
\label{table:simulationpara_ergodiccap}
\end{table}

\begin{figure}
  \centering
  \includegraphics[width=0.48\textwidth]{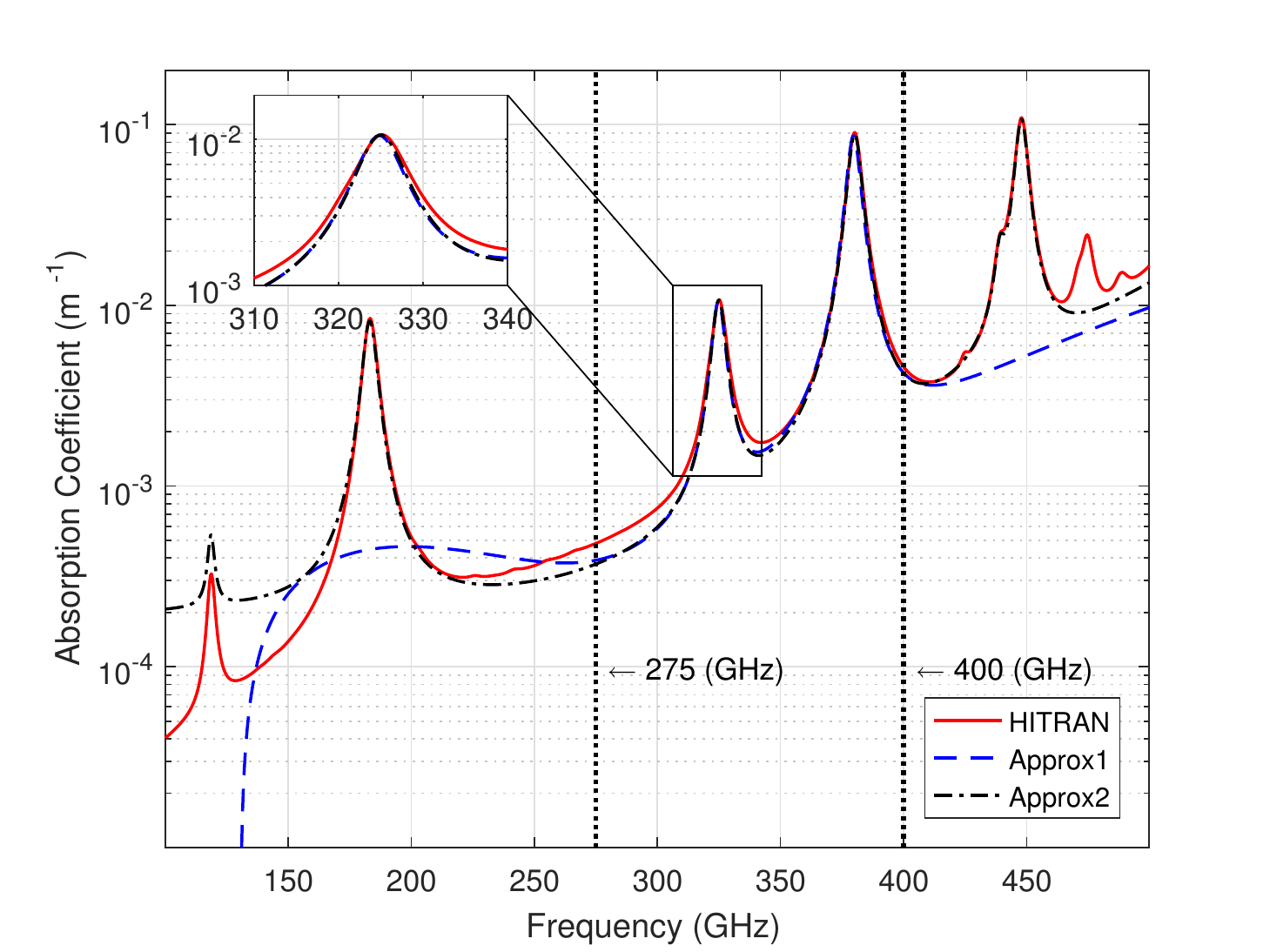}
 \caption{Comparison between the three methods for calculating absorption coefficients.}
  \label{fig:compare_hitran_approx1_approx2}
\end{figure}

\begin{figure}
  \centering
  \includegraphics[width=0.48\textwidth]{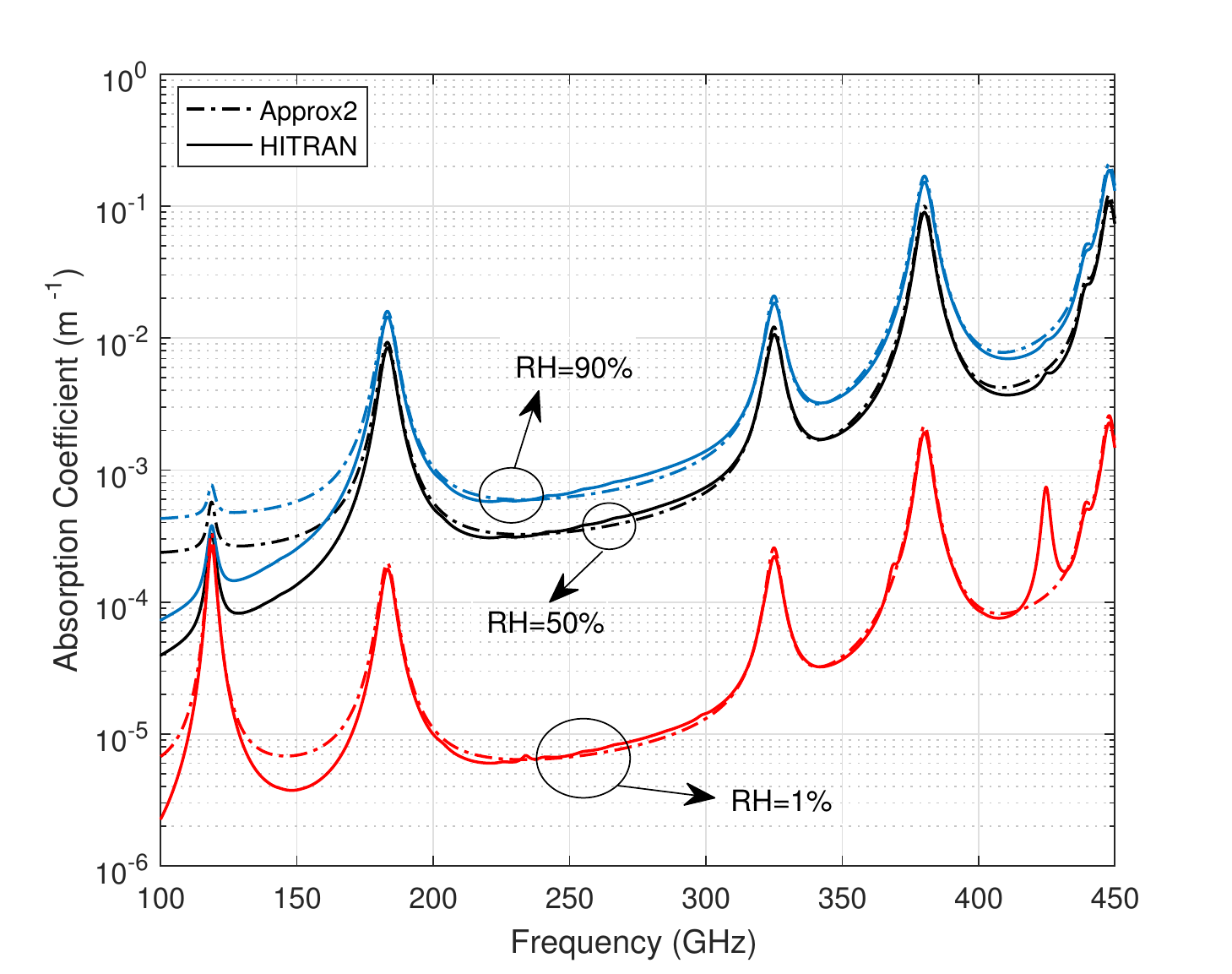}
 \caption{Absorption coefficient for different relative humidity levels: HITRAN (solid) and Approx2 (dash-dot).}
  \label{fig:compare_hit_approx2_diff_RelativeHumidity}
\end{figure}

\begin{figure}
  \centering
  \includegraphics[width=0.5\textwidth]{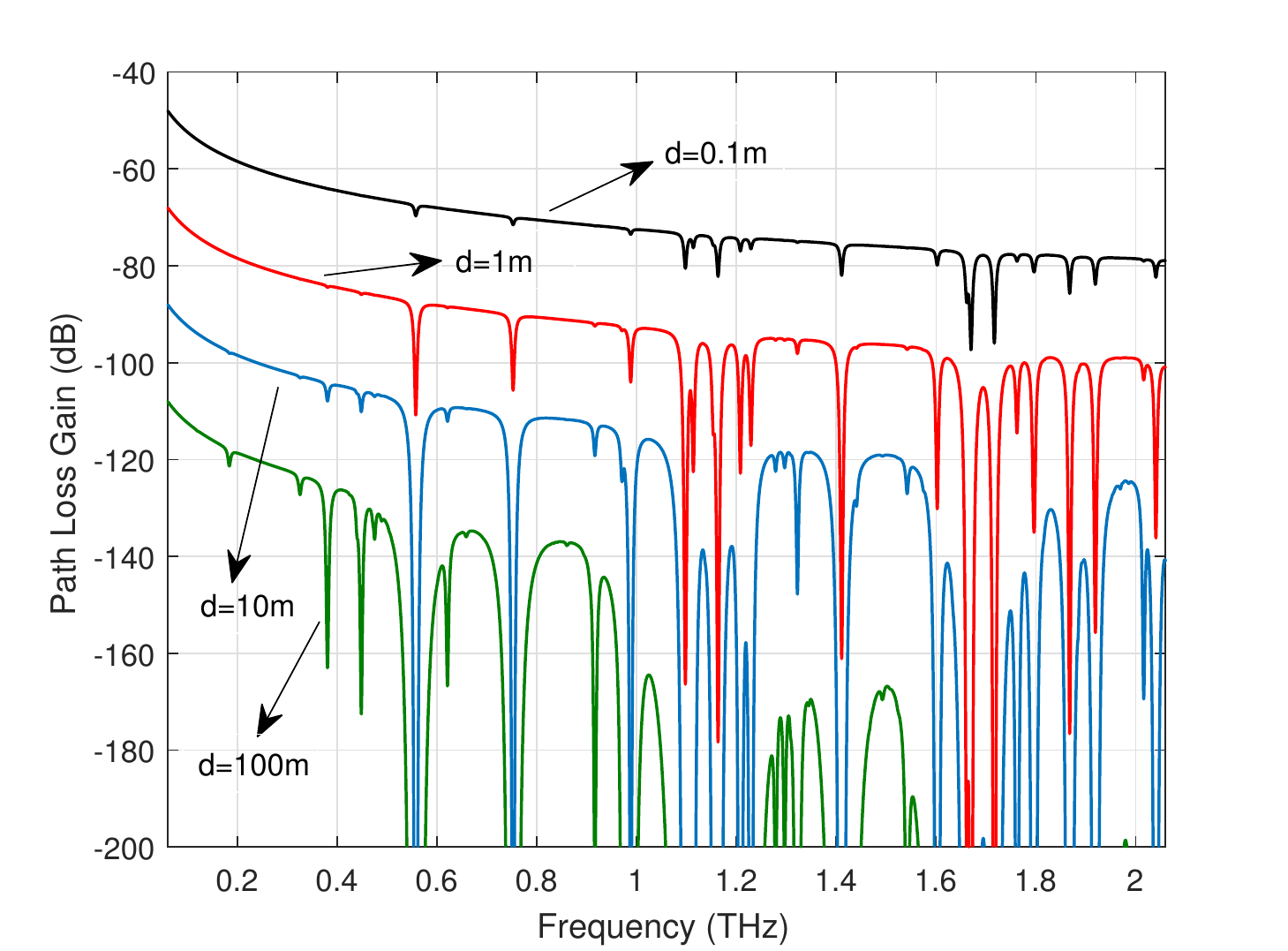}
 \caption{Path loss at different communication distances with HITRAN-based molecular absorption.}
  \label{fig:compare_pathloss_diff_d}
\end{figure}

A comparison between the three molecular absorption models is illustrated Fig. \ref{fig:compare_hitran_approx1_approx2}. The HITRAN-based model shows multiple absorption peaks, such as at $\unit[325]{GHz}$, $\unit[380]{GHz}$, and $\unit[439]{GHz}$. In the sub-THz range, the approximated absorption peaks, and the transmission windows between peaks, greatly match those of the HITRAN-based model.
Fig. \ref{fig:compare_hit_approx2_diff_RelativeHumidity} shows the effect of RH on the molecular absorption coefficient. Increasing the relative humidity increases the absorption coefficient, which in turn increases the total path loss; the simulation results are also generated using both exact and approximate molecular absorption computations, further confirming the robustness of approximations in the sub-THz band. The distance-dependent path loss is illustrated in Fig. \ref{fig:compare_pathloss_diff_d}, where we plot the total path loss, i.e., the spreading and the molecular losses, as a function of frequency. As expected, increasing the communication distance results in severe losses. Moreover, three spectral windows are noted between path loss peaks below $\unit[1]{THz}$; at medium ranges and frequencies higher than $\unit[1]{THz}$, the spectrum gets even more fragmented, where the window widths depend on both the center frequency and the communication distance.

\begin{figure}
\centering
\includegraphics[width=0.5\textwidth]{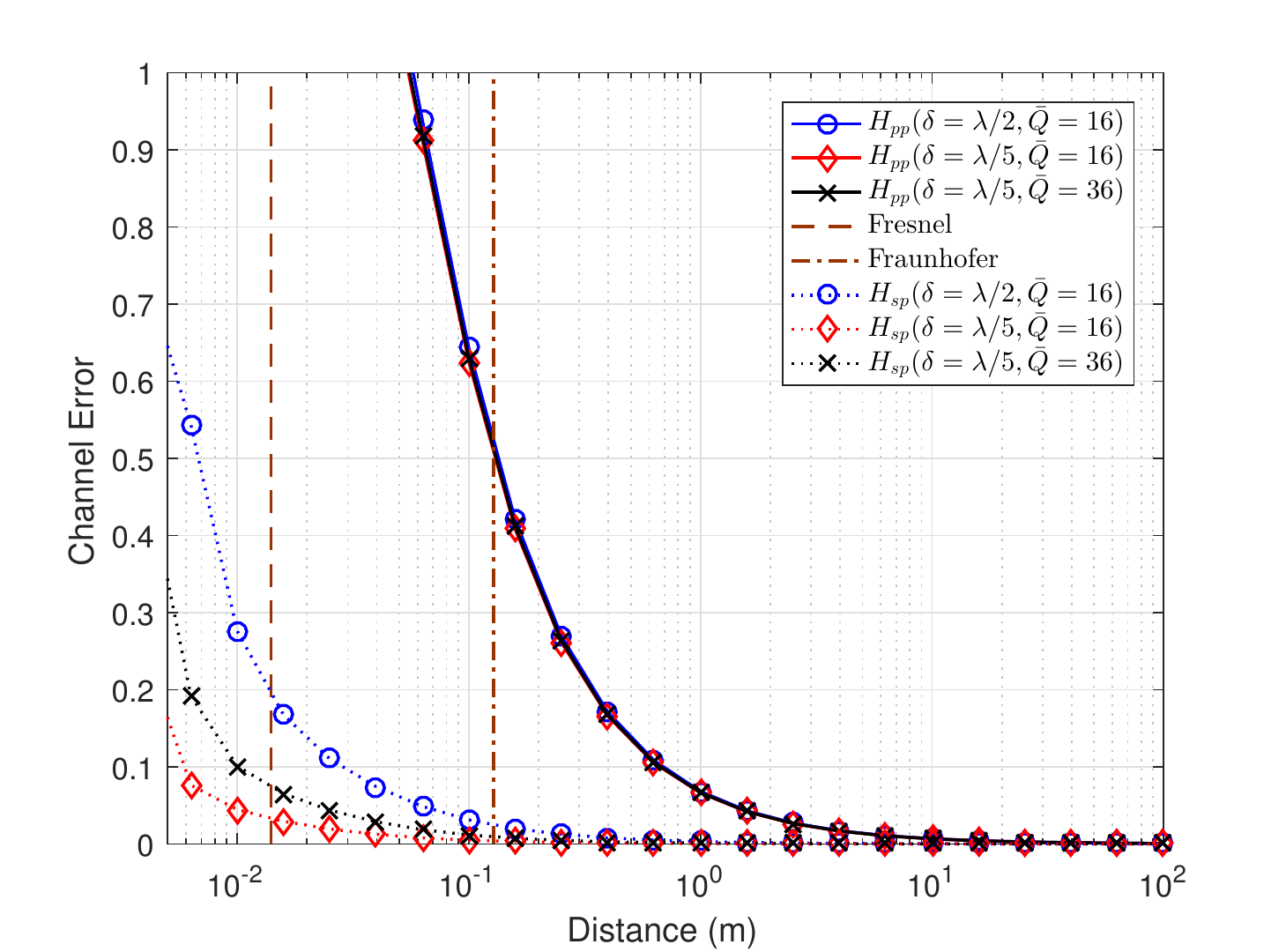}
\caption{Channel realization error between SWM and PWM for different values of $\delta$ with $Q^{(\mathrm{t})}\!=\!Q^{(\mathrm{r})}\!=16$, $\! \bar{Q}^{(\mathrm{t})}\!=\! \bar{Q}^{(\mathrm{r})}\!=\!16/36$, $\Delta=2\lambda_c$, $[\dot{\alpha}^{(\mathrm{t})},\dot{\beta}^{(\mathrm{t})},\dot{\gamma}^{(\mathrm{t})}]=[\frac{\pi}{4},\frac{\pi}{6},0]$, and $[\dot{\alpha}^{(\mathrm{r})},\dot{\beta}^{(\mathrm{r})},\dot{\gamma}^{(\mathrm{r})}]=[\frac{7\pi}{8},0,0]$.}
\label{fig:chanel_error}
\end{figure}

We compare different channel models with the benchmark channel $\mbf{H_{ss}}$. From Fig.~\ref{fig:chanel_error}, we notice that for a fixed array size $D$ in the PWM, the channel errors for different SA sizes are almost the same. However, when SWM is used, the channel error changes depending on the SA size and AE spacing; our proposed model ($\mbf{H_{sp}}$) has much lower errors than the PWM model ($\mbf{H_{pp}}$) and maintains a low computational complexity compared to the SWM (at both SA and AE levels).

\begin{figure}[htb]
\begin{minipage}[b]{0.99\linewidth}
  \centering
  \centerline{\includegraphics[width = 0.8\linewidth]{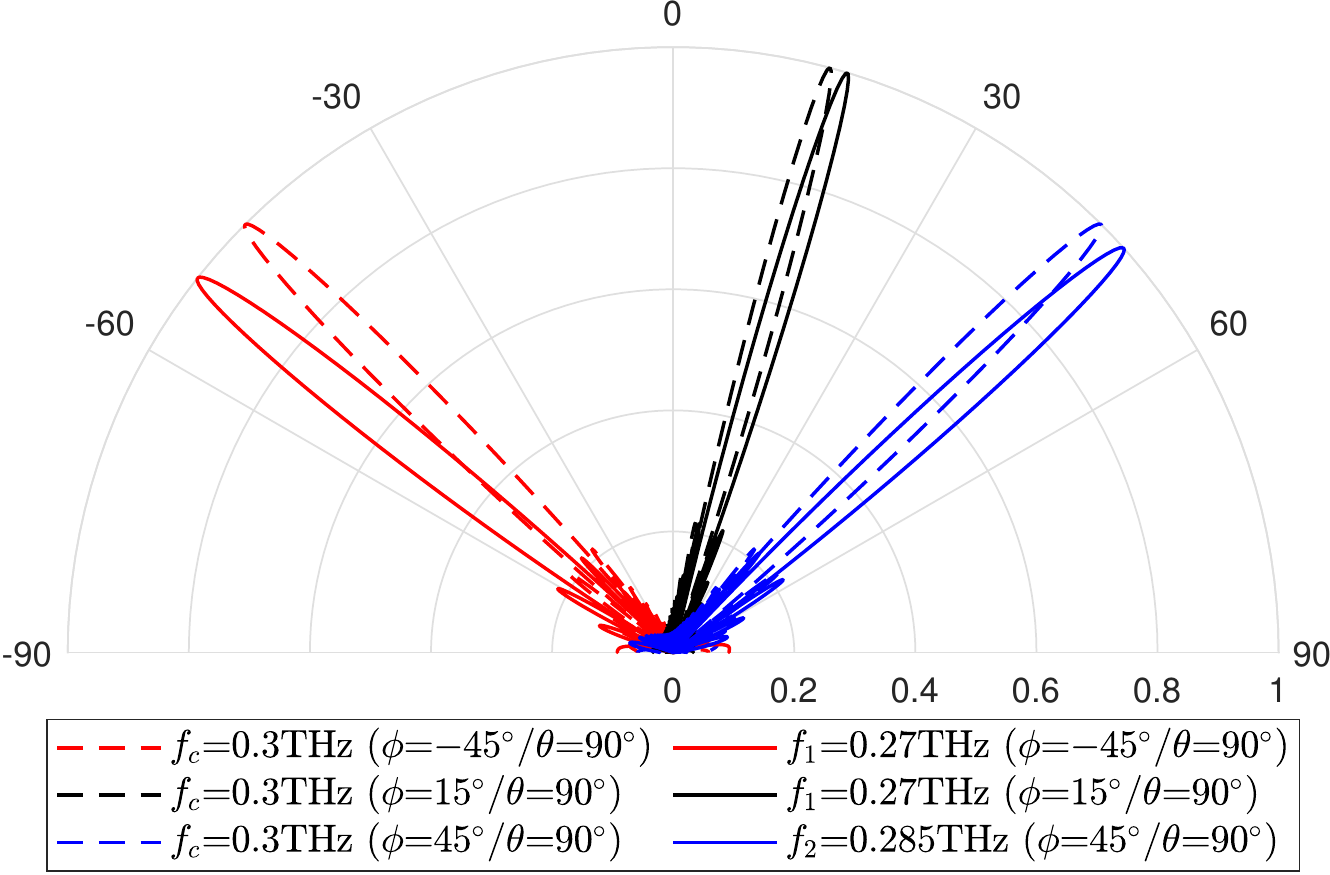}}
  \centerline{(a) Fixed elevation} \medskip
\end{minipage}
\hfill
\begin{minipage}[b]{0.99\linewidth}
  \centering
  \centerline{\includegraphics[width = 0.8\linewidth]{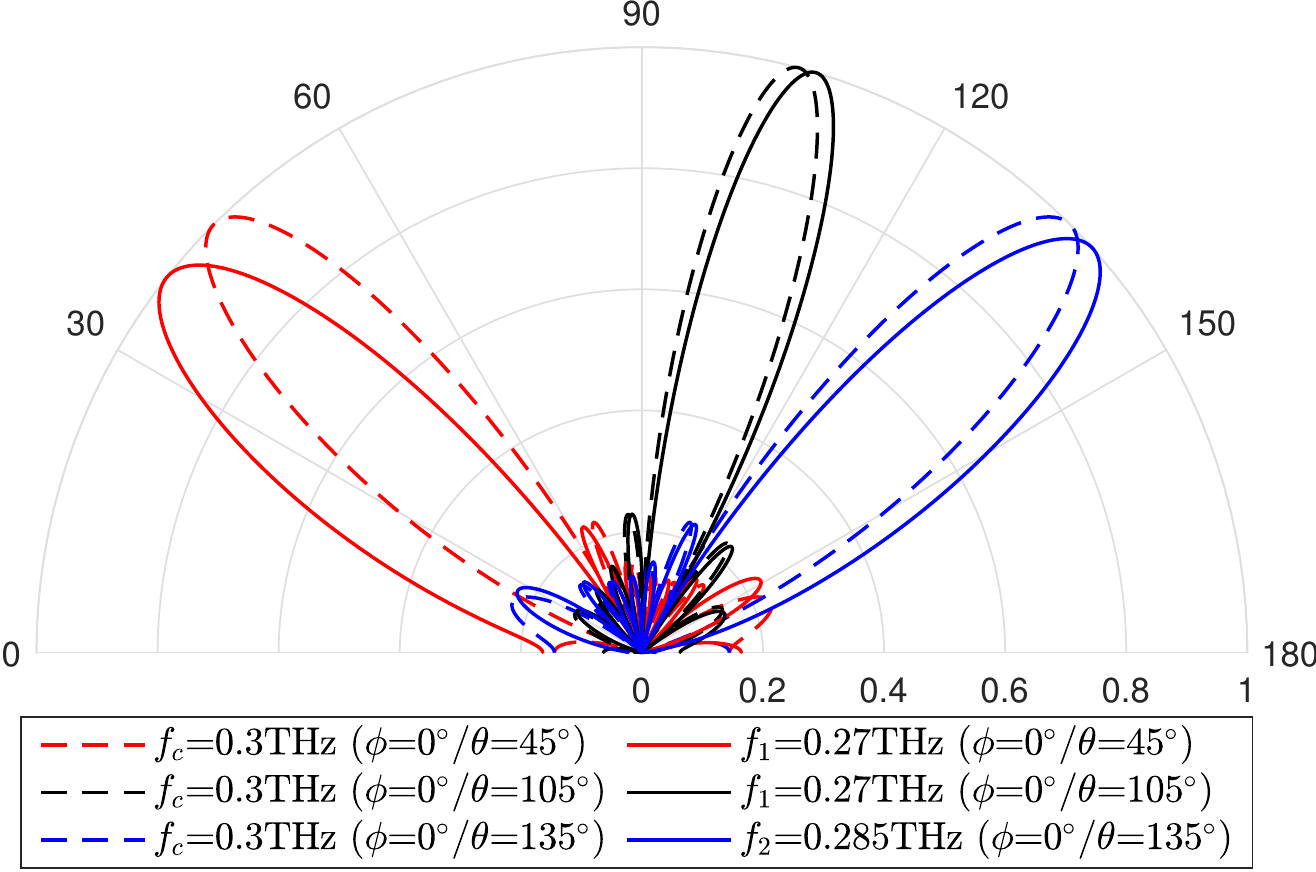}}
\centerline{(b) Fixed azimuth}\medskip
\end{minipage}
\caption{Visualization of beam split effect with $f_1 \!=\!\unit[0.27]{THz}$, $f_2 \!=\!\unit[0.285]{THz}$, and $f_c \!=\!\unit[0.3]{THz}$ on a $8\times32$ UPA with (a) fixed elevation angle $\theta$ and (b) fixed Rx azimuth angle $\phi$.}
\label{fig:beamsplit_2d_upa}
\end{figure}

The beam split effect is further illustrated (as normalized array patterns) in Fig.~\ref{fig:beamsplit_2d_upa} for different beamforming angles and carrier frequencies using an $8\times32$ UPA. In Fig.~\ref{fig:beamsplit_2d_upa} (a), the target elevation is fixed at $90^\circ$, while its azimuth is fixed at $0^\circ$ in Fig.~\ref{fig:beamsplit_2d_upa} (b).
The array gain loss increases when the carrier frequency ($f_k$) deviates from the central frequency ($f_c$). The figure also shows that gain is also affected by the beamforming angle and array size. Narrower beams lead to more severe signal degradation, which should be considered in wide-band systems.

\begin{figure}
  \centering
  \includegraphics[width=0.5\textwidth]{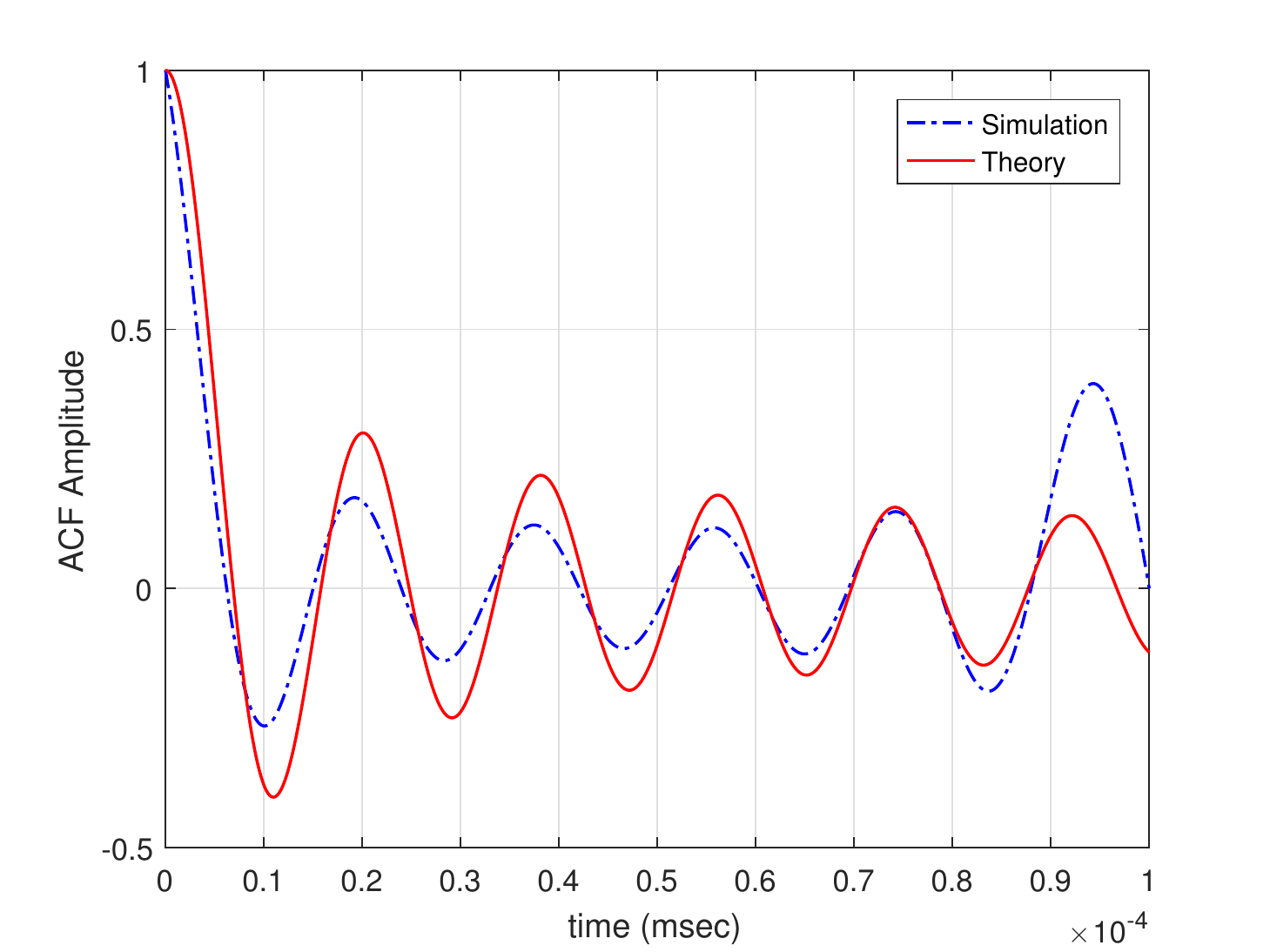}
 \caption{Simulation of TV channel response, at $f_c=\unit[0.5]{THz}$, bandwidth of $B=\unit[1]{GHz}$, and velocity $\vartheta= \unit[120]{km/hr}$.}
 \label{fig:acf_thz_tv_mp}
\end{figure}

Finally, in Fig. \ref{fig:acf_thz_tv_mp}, we plot the simulated ACFs for a TV channel; which closely matches the theoretically predicted one.
\section{Conclusion}
\label{sec:conclusions}

This paper introduces TeraMIMO, a first-of-its-kind comprehensive statistical 3D end-to-end channel simulator for UM-MIMO wideband THz channels. TeraMIMO captures most of the THz channel's peculiarities, such as beam split, misalignment, spherical and planar wave models, and phase uncertainties. The results closely match existing measurements and theoretical bounds available in the literature. This makes TeraMIMO a handy tool for researchers endeavoring in the field of THz communications.

As future work, we aim to incorporate various candidate THz MC schemes, ranging from multiple SCs to CP-OFDM. We also plan to extend TeraMIMO to account for the system-level blockage and shadowing effects, as well as the presence of other infrastructure enablers, such as intelligent reflecting surfaces (IRSs). The extension to multi-user scenarios is also important to make TeraMIMO more scalable and capable of generating network simulations for next-generation wireless standards. Furthermore, as THz device models mature, we plan to incorporate imperfections for both the Tx and Rx, such as the effects of phase noise (PHN), in-phase and quadrature imbalance (IQI), and amplifier non-linearities.

\begin{table}
\footnotesize
\centering
\caption{Summary of Latin Notations}
\begin{tabular} { |c||c|}
 \hline
 Variable & Description \\ [0.5ex] 
 \hline\hline
 $\mbf{a}^{(\mathrm{t})}/\mbf{a}^{(\mathrm{r})}$ & Tx/Rx beamsteering vector\\
 \hline
 $\hat{\mbf{a}}^{(\mathrm{t})}/\hat{\mbf{a}}^{(\mathrm{r})}$ & Tx/Rx beamforming vector\\
 \hline
 $b_{p}$ & misalignment gain\\
 \hline
 $B$ & system bandwidth\\
 \hline
 $B_{\coh}$ & coherence bandwidth\\
 \hline
 $B_{\mathrm{fr}}$ & fractional bandwidth\\ \hline
 $B_{\sig}$ & signal bandwidth\\
 \hline
 $B_{\subb}$ & sub-band bandwidth\\
 \hline
 $c_0$ & speed of light in vacuum\\
 \hline
 C & ergodic capacity\\
 \hline
 $\mbf{C}_{\BBand}[k]$ & \begin{tabular} {c} combining matrices at \\ $\nth{k}$ subcarrier\end{tabular}\\
 \hline
 $\mbf{C}_{\RF}$ & analog RF combiner matrices\\
 \hline
 $d$ & communication distance\\
 \hline
 $\mathrm{dirac}(\cdot)$ & dirac function\\
 \hline
 $D$ & maximum overall antenna dimension\\
 \hline
 $E_L^{i}$ & \begin{tabular}{c} lower state energy of absorbing\\species transition\end{tabular}\\
 \hline
 $f$ & frequency\\
 \hline
 $f_c$ & center frequency\\
 \hline
 $f_{c0}^{(i,g)}$ & \begin{tabular}{c} resonant frequency of $(i,g)$ at\\ zero-pressure position\end{tabular} \\
 \hline
 $f_{c}^{(i,g)}$&\begin{tabular}{c} resonant frequency of the isotopologue\\ $i$ of gas $g$\end{tabular}\\
 \hline
 $f_{\maxx}$ & maximum Doppler shift\\
 \hline
 $f_k$ & $\nth{k}$ subcarrier center frequency\\
 \hline
 $f_{n_{\subb}}$ & $\nth{n_{\subb}}$ sub-band center frequency\\
 \hline 
 g & gas\\
 \hline
 $G^{(\mathrm{t})}/G^{(\mathrm{r})}$ & Tx/Rx antenna gain\\
 \hline
 $\hbar$ & Planck constant\\
 \hline
 $\mbf{H}$ & UM-MIMO channel matrix\\
 \hline
 $\hat{\mbf{H}}$ & effective baseband UM-MIMO channel matrix\\
 \hline
 i & isotopologue of gas $g$ \\
 \hline
 $J_0$ & Bessel function\\
 \hline
 $K$ & number of subcarriers\\
 \hline
 $K_B$ & Boltzmann constant\\
 \hline
 $M^{(\mathrm{t})}/N^{(\mathrm{t})}$ & number of Tx SA Rows/Columns\\
 \hline
 $M^{(\mathrm{r})}/N^{(\mathrm{r})}$ & number of Rx SA Rows/Columns\\
 \hline
 $\bar{M}^{(\mathrm{t})}/\bar{N}^{(\mathrm{t})}$ & \begin{tabular} {c} number of Tx AEs Rows/Columns \\per SA\end{tabular}\\
 \hline
 $\bar{M}^{(\mathrm{r})}/\bar{N}^{(\mathrm{r})}$ & \begin{tabular} {c} number of Rx AEs Rows/Columns \\per SA\end{tabular}\\
 \hline
 $\mbf{M}^{(\mathrm{t})}/\mbf{M}^{(\mathrm{r})}$ & Tx/Rx mutual coupling matrices\\
 \hline
 $\mbf{n}[k]$ & noise vector at $\nth{k}$ subcarrier\\
 \hline
 $N_0$ & noise power\\
 \hline
 $N_A$ & Avogadro constant\\
 \hline
 $N_{\clu}$ & number of clusters\\
 \hline
 $N_{\ray}$ & number of rays\\
 \hline
 $N_{\mathrm{S}}$ & number of transmitted blocks\\
 \hline
 $N_{\subb}$ & number of sub-bands in $B_{\subb}$\\
 \hline
\end{tabular}
\label{table:NotationsTableLatin}
\end{table}

\begin{table}
\footnotesize
\centering
\caption{Summary of Latin Notations}
\begin{tabular} { |c||c|}
 \hline
 Variable & Description \\ [0.5ex] 
 \hline\hline
 $P$ & system pressure\\
 \hline
 $P_0$ & reference pressure\\
 \hline
 $\tilde{\mbf{p}}_{\cent}^{(\mathrm{t})}/\tilde{\mbf{p}}_{\cent}^{(\mathrm{r})}$ & centers of Tx/Rx AoSA (global systems)\\
 \hline
 $\mbf{p}_q^{(\mathrm{t})}/\mbf{p}_q^{(\mathrm{r})}$ & local position of $\nth{q}$ Tx/Rx SA\\
 \hline
 $\tilde{\mbf{p}}_q^{(\mathrm{t})}/\tilde{\mbf{p}}_q^{(\mathrm{r})}$ & global position of $\nth{q}$ Tx/Rx SA\\
 \hline
 $\mbf{p}_{q,\bar{q}}^{(\mathrm{t})}/\mbf{p}_{q,\bar{q}}^{(\mathrm{r})}$ & \begin{tabular} {c} local position of Tx/Rx $\nth{\bar{q}}$ AE,\\ of $\nth{q}$ SA \end{tabular}\\
 \hline
 $\tilde{\mbf{p}}_{q,\bar{q}}^{(\mathrm{t})}/\tilde{\mbf{p}}_{q,\bar{q}}^{(\mathrm{r})}$ & \begin{tabular} {c} global position of Tx/Rx $\nth{\bar{q}}$ AE,\\ of $\nth{q}$ SA \end{tabular}\\
 \hline
 $P_{\mathrm{S}}$ & total average transmit power\\
 \hline
 $P_{\mathrm{Tx}}$ & transmit power\\
 \hline
 $P^{\ast}_{\omega}(T,P)$ & saturated water vapor partial pressure\\
 \hline
 $Q^{(\mathrm{t})}/Q^{(\mathrm{r})}$ & total number of Tx/Rx SAs\\
 \hline
 $\bar{Q}^{(\mathrm{t})}/\bar{Q}^{(\mathrm{r})}$ & total number of Tx/Rx AEs per SA\\
 \hline
 $R$ & gas constant\\
 \hline
 $\mbf{R}$ & rotation matrix\\
 \hline
 $\mbf{s}[k]$ & data symbol at $\nth{k}$ subcarrier\\
 \hline
 $S^{(i,g)}(T)$ & \begin{tabular} {c} line intensity of the isotopologue $i$ of\\ gas $g$\end{tabular}\\
 \hline
 $S_{0}^{(i,g)}$ & \begin{tabular} {c} line intensity of the isotopologue $i$ of\\ gas $g$ at reference temperature\end{tabular}\\
 \hline
 $t$ & time\\
 \hline
 $t_{c,\ell}^{\NLoS}$ & \begin{tabular}{c} time of arrival of the $\nth{\ell}$ ray \\ within the $\nth{c}$ cluster\end{tabular} \\
 \hline
 $\mbf{t}/\tilde{\mbf{t}}$ & local/global unit direction vector\\
 \hline
 ${\mbf{t}}^{(\mathrm{r},\mathrm{t})}/\tilde{\mbf{t}}^{(\mathrm{r},\mathrm{t})}$ & local/global AoD direction \\
 \hline
 ${\mbf{t}}^{(\mathrm{t},\mathrm{r})}/\tilde{\mbf{t}}^{(\mathrm{t},\mathrm{r})}$ & local/global AoA direction \\
 \hline
 $T$ & system temperature\\
 \hline
 $T_0$ & reference temperature\\
 \hline
 $T_c^{\NLoS}$ & time of arrival of the $\nth{c}$ cluster\\
 \hline
 $T_{\coh}$ & coherence time\\
 \hline
 $T_s$ & sampling time \\
 \hline
 $T_{\mathrm{STP}}$ & temperature at standard pressure\\
 \hline
 $T_{\sym}$ & symbol duration\\
 \hline
 U & \begin{tabular} {c} maximum number of resolvable\\ MP delays\end{tabular}\\
 \hline
 $w_1/w_2$ & weights of GMM distribution\\
 \hline
 $w_t(d)$ & \begin{tabular} {c} maximum Tx antenna beam radius at\\ distance $d$\end{tabular}\\
 \hline
 $\mbf{W}_{\BBand}[k]$ & \begin{tabular} {c c} precoding matrices at \\ $\nth{k}$ subcarrier\end{tabular}\\
 \hline
 $\mbf{W}_{\RF}$ & analog RF beamformer matrices\\
 \hline
 $W_{\mathrm{r}}$ & circular effective area of Rx antenna\\
 \hline
 $\mbf{x}[k]$ & transmitted signal at $\nth{k}$ subcarrier\\
 \hline
 $\mbf{y}[k]$ & received signal at $\nth{k}$ subcarrier\\
 \hline
\end{tabular}
\label{table:NotationsTableLatin2}
\end{table}

\begin{table}
\footnotesize
\centering
\caption{Summary of Greek Notations}
\begin{tabular} { |c||c|}
 \hline
 Variable & Description \\ [0.5ex] 
 \hline\hline
 $\alpha^{\LoS}$ & path gain of LoS component\\
 \hline
 $\alpha_0^{(\mathrm{air})}$ & broadening half-widths of air\\
 \hline
 $\alpha_0^{(i,g)}$ & broadening half-widths of $(i,g)$\\
 \hline
 $\alpha^{\NLoS}_{c,\ell}$ & path gain of the $\nth{\ell}$ ray in the $\nth{c}$ cluster\\
 \hline
 $\alpha_L^{(i,g)}$ & Lorentz half-width of $(i,g)$\\
 \hline
 $\dot{\alpha},\dot{\beta},\dot{\gamma}$ & rotation angles around $\mathrm{ZYX}$-axis\\
 \hline
 $\mathcal{A}_{eq}^{(\mathrm{t})}/\mathcal{A}_{eq}^{(\mathrm{r})}$ & Tx/Rx equivalent array response\\
 \hline
 $\beta_{c,\ell}$ & phase shift of the $\nth{\ell}$ ray in the $\nth{c}$ cluster\\
 \hline
 $\gamma$ & path loss exponent\\
 \hline
 $\bar{\gamma}$ & radius of circular effective area Rx antenna\\
 \hline
 $\Gamma/\dot{\Gamma}$ & cluster/ray decay factor\\
 \hline
 $\delta_m^{(\mathrm{t})}/\delta_n^{(\mathrm{t})}$ & \begin{tabular}{c} distance between centers of two Tx\\ adjacent AEs\end{tabular}\\
 \hline
 $\delta_m^{(\mathrm{r})}/\delta_n^{(\mathrm{r})}$ & \begin{tabular}{c} distance between centers of two Rx\\ adjacent AEs\end{tabular}\\
 \hline
 $\Delta_m^{(\mathrm{t})}/\Delta_n^{(\mathrm{t})}$ & \begin{tabular}{c} distance between centers of two Tx\\ adjacent SAs\end{tabular}\\
 \hline
 $\Delta_m^{(\mathrm{r})}/\Delta_n^{(\mathrm{r})}$ & \begin{tabular}{c} distance between centers of two Rx\\ adjacent SAs\end{tabular}\\
 \hline
 $\Delta t$ & time separation\\
 \hline
 $\zeta$ & relative humidity\\
 \hline
 $\theta^{(\mathrm{t})}/\theta^{(\mathrm{r})}$ & Tx/Rx elevation angle\\
 \hline
 $\theta_{0}^{(\mathrm{t})}/\theta_{0}^{(\mathrm{r})}$ & Tx/Rx target elevation angle\\
 \hline
 $\theta^{(\mathrm{t})}_{c,\ell}/\theta^{(\mathrm{r})}_{c,\ell}$ &\begin{tabular}{c c} total elevation angle of the $\nth{\ell}$ ray in the \\$\nth{c}$ cluster at Tx/Rx\end{tabular}\\
 \hline
 $\hat{\theta}_{c,\ell}^{(\mathrm{t})}/\hat{\theta}_{c,\ell}^{(\mathrm{r})}$ & AoD/AoA ray elevation angle\\
 \hline
 $\Theta_{c}^{(\mathrm{t})}$/$\Theta_{c}^{(\mathrm{r})}$ & cluster AoD/AoA elevation angle\\
 \hline
 $\vartheta$ & velocity of BS/UE \\
 \hline
 $\iota$ & temperature broadening coefficient\\
 \hline
 $\lambda$ & free-space wavelength\\
 \hline
 $\lambda_c$ & wavelength at $f_c$\\
 \hline
 $\lambda_k$ & wavelength at $f_k$\\
 \hline
 $\lambda_{\spp}$ & SPP wavelength\\
 \hline
 $\Lambda/\dot{\Lambda}$ & cluster/ray arrival rate\\
 \hline
 ${\mu_{\htwoo}}$ & volume mixing ratio of water vapor\\
 \hline
 $\nu$ & Doppler\\
 \hline
 $\xi^{(i,g)}$ & mixing ratio of the isotopologue $i$ of gas $g$\\
 \hline
 $\Xi(\cdot,\cdot)$ & ACF of TV channel\\
 \hline
 $\rho$ & Tx circular beam footprint at a distance $d$\\
 \hline
\end{tabular}
\label{table:NotationsTableGreek}
\end{table}

\begin{table}
\footnotesize
\centering
\caption{Summary of Greek Notations}
\begin{tabular} { |c||c|}
 \hline
 Variable & Description \\ [0.5ex] 
 \hline\hline
 $\sigma_n^2$ & noise power\\
 \hline
 $\sigma^2_1/\sigma^2_2$& variances of GMM distribution\\
 \hline
 $\varsigma^{(i,g)}$ & linear pressure shift of $(i,g)$\\
 \hline
 $\tau$ & delay\\
 \hline 
 $\tau^{\LoS}$ & time delay of LoS component\\
 \hline
 $\tau^{\NLoS}_{c,\ell}$ & time delay of the $\nth{\ell}$ ray in the $\nth{c}$ cluster\\
 \hline
 $\tau_{\rms}$ & RMS delay spread\\
 \hline
 $\phi^{(\mathrm{t})}/\phi^{(\mathrm{r})}$ & Tx/Rx azimuth angle\\ 
 \hline
 $\phi_{0}^{(\mathrm{t})}/\phi_{0}^{(\mathrm{r})}$ & Tx/Rx target azimuth angle\\ 
 \hline
 $\phi^{(\mathrm{t})}_{c,\ell}/\phi^{(\mathrm{r})}_{c,\ell}$ &\begin{tabular}{c} total azimuth angle of the $\nth{\ell}$ ray in the \\$\nth{c}$ cluster at Tx/Rx\end{tabular}\\
 \hline
 $\varphi_{c,\ell}^{(\mathrm{t})}/\varphi_{c,\ell}^{(\mathrm{r})}$ & AoD/AoA ray azimuth angle\\
 \hline
 $\Phi_{c}^{(\mathrm{t})}$/$\Phi_{c}^{(\mathrm{r})}$ & AoD/AoA cluster azimuth angle\\
 \hline
 $\mbf{\Phi}^{(\mathrm{t})}/\mbf{\Phi}^{(\mathrm{r})}$ & LoS angle of departure/arrival vector\\
 \hline
 $\mbf{\Phi}_{0}^{(\mathrm{t})}/\mbf{\Phi}_{0}^{(\mathrm{r})}$ & Target angle of departure/arrival vector\\
 \hline
 $\mbf{\Phi}_{c,\ell}^{(\mathrm{t})}/\mbf{\Phi}_{c,\ell}^{(\mathrm{r})}$ & \begin{tabular}{c} angle of departure/arrival vector of the\\ $\nth{\ell}$ ray in the $\nth{c}$ cluster\end{tabular}\\
 \hline
 $\psi_\mathrm{A}$ & solid angle\\
 \hline
 $\psi_\phi,\psi_\theta$ & azimuth/elevation plane HPBW\\
 \hline
 $\Psi_{\bar{q}}$ & $\nth{\bar{q}}$ AE phase shifts\\
 \hline
 $\hat{\Psi}_{\bar{q}}$ & AE spatial direction \\
 \hline
 $\CC$ & set of complex numbers\\
 \hline
 $\mathcal{K}(f)$ & molecular absorption coefficient\\
 \hline
 $\mathcal{S}(\nu)$ & Doppler spectrum\\
 \hline
 $\mathcal{X}$ & modulation constellation\\
 \hline
\end{tabular}
\label{table:NotationsTableGreek2}
\end{table}


\newpage

\section*{Biographies}
\footnotesize

\textbf{Simon Tarboush} received his B.E. degree in Communication Engineering from Damascus, Syria, in 2014, and his MSc. degree (with honors) in Radio and Mobile Communication Systems in 2019. His research interests lie in the areas of waveform design, digital communication, and signal processing for modern wireless communications.

\textbf{Hadi Sarieddeen} (S'13-M'18) received his B.E. degree (summa cum laude) in Computer and Communications Engineering from Notre Dame University-Louaize (NDU), Lebanon, in 2013, and his Ph.D. degree in Electrical and Computer Engineering from the American University of Beirut (AUB), Beirut, Lebanon, in 2018. He is currently a postdoctoral research fellow at KAUST. His research interests are in the areas of communication theory and signal processing for wireless communications.

\textbf{Hui Chen} (S'18-M'21) received his B.S. degree in Electrical Engineering from Beijing Forestry University, Beijing, China, in 2013, the M.S. degree in Computer Application Technology from the University of Chinese Academy of Sciences (UCAS), Beijing, China, in 2016, and the Ph.D. degree in Electrical and Computer Engineering from KAUST, in 2021. He is currently a postdoctoral researcher at the Chalmers University of Technology, Sweden. His research interests include localization \& tracking, stochastic optimization, and machine learning for signal processing.

\textbf{Mohamed Habib Loukil} received his B.E. degree in Signals and Systems Engineering from École Polytechnique de Tunisie (EPT), Tunisia, in 2019, and his Master's degree in Electrical and Computer Engineering from KAUST, in 2021. He is currently a Ph.D. student at North Carolina State University, United States. His research interests are in the areas of communication theory and signal processing for wireless communications.

\textbf{Hakim Jemaa} received his B.E. degree in Multidisciplinary Engineering from Tunisia Polytechnic School, Tunisia, in 2020. He is currently an M.S./Ph.D. student at the Communication Theory Lab at KAUST. His research interests are in the area of signal processing for wireless communications.

\textbf{Mohamed-Slim Alouini} (S'94-M'98-SM'03-F'09) was born in Tunis, Tunisia. He received his Ph.D. degree in Electrical Engineering from Caltech, CA, USA, in 1998. He served as a faculty member at the University of Minnesota, Minneapolis, MN, USA, then in Texas A\&M University at Qatar, Doha, Qatar, before joining KAUST as a Professor of Electrical Engineering in 2009. His current research interests include the modeling, design, and performance analysis of wireless communications systems.

\textbf{Tareq Y. Al-Naffouri} (M'10-SM'18) Tareq Al-Naffouri received his B.S. degrees in Mathematics and Electrical Engineering (with first honors) from KFUPM, Saudi Arabia, his M.S. degree in Electrical Engineering from Georgia Tech, Atlanta, in 1998, and his Ph.D. degree in Electrical Engineering from Stanford University, Stanford, CA, in 2004. He is currently a Professor at the Electrical Engineering Department, KAUST. His research interests lie in the areas of sparse, adaptive, and statistical signal processing and their applications, localization, machine learning, and network information theory.

\end{document}